\documentstyle[graphicx,lscape,times]{mn}

\newif\ifAMStwofonts

\ifoldfss
  \ifCUPmtlplainloaded \else
    \NewTextAlphabet{textbfit} {cmbxti10} {}
    \NewTextAlphabet{textbfss} {cmssbx10} {}
    \NewMathAlphabet{mathbfit} {cmbxti10} {} 
    \NewMathAlphabet{mathbfss} {cmssbx10} {} 
  \fi
  \ifAMStwofonts
    \ifCUPmtlplainloaded \else
      \NewSymbolFont{upmath} {eurm10}
      \NewSymbolFont{AMSa} {msam10}
      \NewMathSymbol{\upi}     {0}{upmath}{19}
      \NewMathSymbol{\umu}     {0}{upmath}{16}
      \NewMathSymbol{\upartial}{0}{upmath}{40}
      \NewMathSymbol{\leqslant}{3}{AMSa}{36}
      \NewMathSymbol{\geqslant}{3}{AMSa}{3E}

       \let\le=\leqslant
       
    \fi
  \fi
\fi 

\ifnfssone
  \newmathalphabet{\mathit}
  \addtoversion{normal}{\mathit}{cmr}{m}{it}
  \addtoversion{bold}{\mathit}{cmr}{bx}{it}
  \newmathalphabet{\mathbfit} 
  \addtoversion{normal}{\mathbfit}{cmr}{bx}{it}
  \addtoversion{bold}{\mathbfit}{cmr}{bx}{it}
  \newmathalphabet{\mathbfss} 
  \addtoversion{normal}{\mathbfss}{cmss}{bx}{n}
  \addtoversion{bold}{\mathbfss}{cmss}{bx}{n}
  \ifAMStwofonts
    \ifCUPmtlplainloaded \else
      \UseAMStwoboldmath
      \makeatletter
      \new@mathgroup\upmath@group
      \define@mathgroup\mv@normal\upmath@group{eur}{m}{n}
      \define@mathgroup\mv@bold\upmath@group{eur}{b}{n}
      \edef\UPM{\hexnumber\upmath@group}
      \new@mathgroup\amsa@group
      \define@mathgroup\mv@normal\amsa@group{msa}{m}{n}
      \define@mathgroup\mv@bold\amsa@group{msa}{m}{n}
      \edef\AMSa{\hexnumber\amsa@group}
      \makeatother
      \mathchardef\upi="0\UPM19
      \mathchardef\umu="0\UPM16
      \mathchardef\upartial="0\UPM40
      \mathchardef\leqslant="3\AMSa36
      \mathchardef\geqslant="3\AMSa3E

       \let\le=\leqslant

    \fi
  \fi
\fi 

\ifnfsstwo
  \DeclareMathAlphabet{\mathbfit}{OT1}{cmr}{bx}{it}
  \SetMathAlphabet\mathbfit{bold}{OT1}{cmr}{bx}{it}
  \DeclareMathAlphabet{\mathbfss}{OT1}{cmss}{bx}{n}
  \SetMathAlphabet\mathbfss{bold}{OT1}{cmss}{bx}{n}
  \ifAMStwofonts
    \ifCUPmtlplainloaded \else
      \DeclareSymbolFont{UPM}{U}{eur}{m}{n}
      \SetSymbolFont{UPM}{bold}{U}{eur}{b}{n}
      \DeclareSymbolFont{AMSa}{U}{msa}{m}{n}
      \DeclareMathSymbol{\upi}{0}{UPM}{"19}
      \DeclareMathSymbol{\umu}{0}{UPM}{"16}
      \DeclareMathSymbol{\upartial}{0}{UPM}{"40}
      \DeclareMathSymbol{\leqslant}{3}{AMSa}{"36}
      \DeclareMathSymbol{\geqslant}{3}{AMSa}{"3E}

       \let\le=\leqslant

    \fi
  \fi
\fi 

\ifCUPmtlplainloaded \else
  \ifAMStwofonts \else 
    \def\upi{\pi}
    \def\umu{\mu}
    \def\upartial{\partial}
  \fi
\fi

\title[Galactic globular cluster sub-systems]
  {The properties of Galactic globular cluster sub-systems}
\author[A.~D.~Mackey \& Sidney~van~den~Bergh]
  {A.~D.~Mackey$^1$ and Sidney~van~den~Bergh$^2$\\
  $^1$Institute of Astronomy, University of Cambridge, Madingley Road,
  Cambridge CB3 0HA, UK. \hspace{1mm} E-mail: dmackey@ast.cam.ac.uk \\
  $^2$Dominion Astrophysical Observatory, Herzberg Institute of 
  Astrophysics, National Research Council of Canada, 5071 West Saanich Road, \\ 
  \hspace{5mm}Victoria, British Columbia, V9E 2E7, Canada. 
  \hspace{1mm} E-mail: sidney.vandenbergh@nrc-cnrc.gc.ca}
\date{Accepted --. Received --}
\pagerange{000--000}
\pubyear{2004}

\def\LaTeX{L\kern-.36em\raise.3ex\hbox{a}\kern-.15em
    T\kern-.1667em\lower.7ex\hbox{E}\kern-.125emX}

\begin{document}

\label{firstpage}

\maketitle

\begin{abstract}
In this paper we compare the properties of three sub-systems of
Galactic globular clusters, which are defined according to
metallicity and horizontal branch morphology. We specifically focus on
cluster luminosities, structures, surface brightnesses and ellipticities. 
It is shown that the so-called ``young'' halo (YH) clusters, which
are thought to have formed in external satellite galaxies, exhibit
characteristics which are clearly distinct from those of the ``old'' halo 
(OH) and bulge/disk (BD) clusters, the majority of which are believed
to be Galactic natives. The properties of the YH objects are, in many
respects, similar to those of clusters belonging to a number of present day 
satellite dwarf galaxies. The OH and BD populations have apparently been
strongly modified by destructive tidal forces and shocks in the
inner Galaxy. By comparing the properties of the three cluster
sub-systems, we estimate that the present population of native Galactic
clusters may only represent approximately two-thirds of the original
population. Several clusters with low surface brightnesses are observed
to be highly flattened. We briefly speculate on the possibility that
this ellipticity reflects the intrinsic flattening of dark matter
mini-halos in which these optically dim clusters might be embedded.
Finally, we examine the distribution of clusters on the size ($\log R_{\rm{h}}$)
vs. luminosity ($M_V$) plane. Three objects are seen to fall
well above the sharp upper envelope of the main distribution of clusters
on the size-luminosity plane: $\omega$ Centauri, M54, and NGC 2419. All
three of these objects have previously, and independently, been suggested 
to be the stripped cores of former dwarf galaxies. This suspicion is
strengthened by the additional observation that the massive cluster G1
in M31, plus a number of the most luminous clusters in NGC 5128 also fall
in the same region of the $\log R_{\rm{h}}$ vs. $M_V$ plane. All of the latter
objects have previously been suggested as the stripped cores of now 
defunct dwarf galaxies.
\end{abstract}


\section{Introduction}
\label{s:intro}
Our Galaxy contains $\sim 150$ globular clusters, located at distances
between $500$ pc and $120$ kpc from the Galactic centre. Because they
are generally bright, and probe such a large range of Galactocentric
radii, these objects constitute important tools for both tracing the
properties of our Galaxy in the present epoch and for piecing together
its formation history. They are also interesting objects in their own
right, since they have the potential to provide information on star
and cluster formation processes. Furthermore they provide useful
laboratories for the study of self-gravitating stellar systems.

As a globular cluster evolves dynamically its core contracts
and its envelope expands. However, $N$-body calculations (e.g.,
Spitzer \& Thuan \shortcite{spitzer:72}; Lightman \& Shapiro 
\shortcite{lightman:78}; Aarseth \& Heggie \shortcite{aarseth:98})
show that $R_{\rm{h}}$ -- the radius that contains half of the cluster stars in
projection, generally changes very little over periods as long as ten
relaxation times. This suggests that $R_{\rm{h}}$, like $[$Fe$/$H$]$, may provide
valuable information on physical conditions during the era of globular
cluster formation. However, an important caveat is that metallicity
dependent differences of up to $20$ per cent can occur between the half-light
radii and the half-mass radii of clusters of differing metallicity
\cite{jordan:04}. The reason for this metallicity dependence is that
the masses of the most luminous surviving stars become larger as the
cluster metallicity increases. As a result the density profile of a
metal-rich cluster will be slightly more centrally concentrated than
that of an otherwise similar metal-poor cluster. Nevertheless, the
half-light radii of clusters are of considerable interest because
they are stable over many relaxation times.
   
The work of Mackey \& Gilmore \shortcite{mackey:04} has recently 
re-emphasized the fact that Galactic globular clusters comprise a very 
inhomogeneous class of objects. It is therefore useful to try to identify
sub-groups made up of clusters having similar characteristics. One of
the most successful classification schemes for the Galactic globular
clusters is that introduced over a decade ago (see e.g., Zinn 
\shortcite{zinn:93}; van den Bergh \shortcite{vdb:93}, and references 
therein), in which clusters are grouped according to both their metallicity 
and their horizontal-branch (HB) morphology. These parameters are closely 
related to a number of other cluster properties such as, for example, 
Galactocentric radius and age. Such a division of globular clusters can 
therefore provide us with a great deal of information about the evolutionary 
history of the Galaxy. Recently, Mackey \& Gilmore \shortcite{mackey:04} 
updated the original classification scheme to cover the full Galactic 
globular cluster sample for which information is presently available. By 
comparing the different Galactic sub-groups to the cluster systems in 
nearby dwarf galaxies (i.e., the LMC, SMC, and Fornax and Sagittarius dwarf 
spheroidal galaxies) these authors confirmed that $\sim 30$ per cent of the 
Galactic globular cluster system possesses 
properties consistent with an origin external to the Milky Way. In other 
words, such clusters are most likely to have formed within much smaller 
galaxies (perhaps similar to the Local Group dwarf spheroidals observed 
at the present epoch) which later merged with the Milky Way system. 
Furthermore, Mackey \& Gilmore \shortcite{mackey:04} showed that classifying 
Galactic globular clusters by metallicity and HB morphology leads quite 
naturally to a segregation of clusters by size. These authors used cluster 
core radius $R_c$ as their size diagnostic; however, unlike the half-light 
radius $R_{\rm{h}}$, this parameter is quite sensitive to the dynamically evolving 
state of a globular cluster. It is therefore of interest to extend the 
analysis of Mackey \& Gilmore \shortcite{mackey:04} to additional parameters, 
such as $R_{\rm{h}}$, which ``remember'' more about the initial sizes of clusters, 
and hence about physical conditions in the Galaxy and/or in its progenitors, 
at early epochs.

In the present paper we shall examine the different Galactic
globular cluster sub-systems in terms of cluster structures,
luminosities, and surface brightnesses. Section \ref{s:data} lists the 
relevant data for each of the Galactic globular clusters and presents the
classification of each object according to Mackey \& Gilmore 
\shortcite{mackey:04}. In Section \ref{s:properties} we use these data 
to investigate the collective properties of each cluster sub-system, while 
in Section \ref{s:rhmvplane} we discuss the possible links between half-light 
radius and luminosity both for Galactic globular clusters and for clusters 
belonging to some other nearby Local Group dwarf galaxies. Finally, 
we discuss what these parameters might be able to tell us about the 
origins and evolution of different types of clusters.

\section{Cluster data and classifications}
\label{s:data}
In Table \ref{t:alldata} we list all $150$ known Galactic globular clusters, 
including the handful of clusters that are thought to be associated with the Sagittarius 
dwarf galaxy. The structures of most of these objects have been measured.
For each such cluster we list the half-light radius ($R_{\rm{h}}$) and tidal
radius ($R_{\rm{t}}$). We have adopted the majority of these values from the
online database of Harris \shortcite{harris:96} (2003 update). Since this 
compilation lists $R_{\rm{h}}$ and $R_{\rm{t}}$ as angular diameters, we have converted 
them to parsecs using the cluster distances that are also listed by Harris. 
For a few clusters the database does not have entries for $R_{\rm{h}}$ and $R_{\rm{t}}$. 
We have been able to fill in these missing values for four clusters: IC 1257
\cite{cote:99}, BH 176, Terzan 4, and 1636-283 \cite{webbink:85}. In
addition we have, for HP 1, adopted the measurements of Ortolani, Bica
\& Barbuy \shortcite{ortolani:97}, who used a full colour-magnitude diagram 
to derive a distance less than half that listed by Harris, in good agreement 
with the measurement by Davidge \shortcite{davidge:00}. Also given in Table 
\ref{t:alldata} are the Galactocentric distance $R_{\rm{gc}}$, the integrated 
luminosity $M_V$, and (where available) the ellipticity ($\epsilon = 1 - b/a$, where 
$a$ is the cluster semi-major axis and $b$ the semi-minor axis), 
again all taken from the Harris database.
   
Using the half-light radius and integrated luminosity for each
cluster, we have calculated $I_{\rm{h}}$, the half-light intensity.
This is defined as $I_{\rm{h}} = \frac{L}{2\pi R_{\rm{h}}^2}$, and is measured in 
units of solar luminosities per square parsec ($L_\odot\,\,$pc$^{-2}$). 
It is essentially a measure of the mean surface brightness of a cluster
within $R_{\rm{h}}$, but in terms of absolute units rather than observational units.
Because of the evaporation of stars from dynamically evolving clusters, $I_{\rm{h}}$ 
is more sensitive to evolutionary effects
than is $R_{\rm{h}}$. Nevertheless, this parameter turns out to be useful when
examining cluster classifications. Finally, Table \ref{t:alldata} lists the
metallicity $[$Fe$/$H$]$ and HB morphology index of each cluster, along 
with the formal type classification from Mackey \& Gilmore 
\shortcite{mackey:04}. As noted in Section \ref{s:intro}, this classification 
scheme follows closely those of Zinn \shortcite{zinn:93} and of van den Bergh 
\shortcite{vdb:93} (see also the references therein). To briefly review this 
scheme, the clusters are grouped by metallicity and HB-type\footnote{HB-type 
is parametrized by the dimensionless morphology index $\frac{(B-R)}{(B+V+R)}$
\cite{lee:90,lee:94}, where $B$ is the number of HB stars which lie to 
the blue of the instability strip; $R$ is the number of HB stars falling 
to the red of the instability strip; and $V$ is the number of variable 
HB stars}. The metal rich globular clusters with $[$Fe$/$H$] > -0.8$ all 
have red horizontal branches and are spatially confined to the bulge and 
inner disk of the Galaxy. Clusters in this sub-group are designated 
bulge/disk (BD) objects. In contrast, the metal-poor clusters (with 
$[$Fe$/$H$] \le -0.8$) are seen to have a wide range in HB-type and are 
generally situated in the Galactic halo.

Most of the inner halo clusters exhibit quite a tight relationship
between HB-type and $[$Fe$/$H$]$. On the other hand many of the outer halo
clusters have much redder HB-types at given metallicity (this is
the classical ``second-parameter phenomenon'', where a parameter in
addition to metallicity affects HB morphology). Following Zinn
\shortcite{zinn:93}, the ``old'' halo (OH) and ``young'' halo (YH) 
sub-systems are defined by measuring the offset in HB-type at given 
$[$Fe$/$H$]$ of each cluster from the fiducial line which traces the 
relationship between HB-type and $[$Fe$/$H$]$ for the inner halo 
clusters\footnote{Mackey \& Gilmore \shortcite{mackey:04} used the oldest 
isochrone from the theoretical HB models of Rey et al. \shortcite{rey:01} 
as their fiducial}. Objects for which the offset in HB-type is greater than 
$-0.3$ (i.e., objects which do not lie far to the red of the fiducial line) 
constitute the OH class, while those with offset less than $-0.3$ (i.e., 
clusters which have red HB-type at given $[$Fe$/$H$]$) form the YH class.

The majority of the metallicity and HB-type parameters we list in
Table \ref{t:alldata} are taken from the Harris database, except where 
they were updated by Mackey \& Gilmore \shortcite{mackey:04} (one 
metallicity value and $28$ HB indices). There are five classification 
types listed in Table \ref{t:alldata}: BD for the bulge/disk clusters 
($37$ objects); OH for the old halo clusters ($70$ objects); YH for the 
young halo clusters ($30$ objects); SG for clusters belonging to the 
Sagittarius dwarf ($6$ objects); and UN for clusters whose membership 
type remains unknown due to insufficient or unclassifiable data ($7$ 
objects).

In the following two sections, we consider some of the properties
of these cluster sub-systems which were not considered by Mackey \&
Gilmore, but which nevertheless provide useful insights into the origin
and evolution of objects in each of the different classes of Galactic
globular cluster.

\begin{table*}
\begin{center}
\begin{minipage}{126mm}
\caption{Data and classifications for Galactic globular clusters.}
\begin{tabular}{@{}lccccccccc}
\hline \hline
Cluster & $R_{\rm{gc}}$ & $M_V$ & $R_{\rm{h}}$ & $R_{\rm{t}}$ & $\log\,I_{\rm{h}}$ & $\epsilon$ & $[$Fe$/$H$]$ & HB-index & Class \\
Name & (kpc) & & (pc) & (pc) & ($\rm{L}_{\odot}\,\,$pc$^{-2}$) & ($1-b/a$) & & & \\
\hline
NGC 104  & $7.4$ & $-9.42$ & $3.65$ & $56.10$ & $3.77$ & $0.09$ & $-0.76$ & $-0.99$ & BD \\
NGC 288  & $12.0$ & $-6.74$ & $5.68$ & $33.12$ & $2.32$ & $...$ & $-1.24$ & $0.98$ & OH \\
NGC 362  & $9.4$ & $-8.41$ & $2.00$ & $39.83$ & $3.89$ & $0.01$ & $-1.16$ & $-0.87$ & YH \\
NGC 1261 & $18.2$ & $-7.81$ & $3.58$ & $34.73$ & $3.15$ & $0.07$ & $-1.35$ & $-0.71$ & YH \\
Pal. 1   & $17.0$ & $-2.47$ & $2.16$ & $28.41$ & $1.45$ & $0.22$ & $-0.60$ & $-1.00$ & BD \\
AM-1     & $123.2$ & $-4.71$ & $17.73$ & $68.08$ & $0.52$ & $...$ & $-1.80$ & $-0.93$ & YH \\
Eridanus & $95.2$ & $-5.14$ & $10.50$ & $83.17$ & $1.14$ & $...$ & $-1.46$ & $-1.00$ & YH \\
Pal. 2   & $35.4$ & $-8.01$ & $5.38$ & $54.27$ & $2.87$ & $0.05$ & $-1.30$ & $-0.10$ & YH \\
NGC 1851 & $16.7$ & $-8.33$ & $1.83$ & $41.18$ & $3.94$ & $0.05$ & $-1.22$ & $-0.32$ & OH \\
NGC 1904 & $18.8$ & $-7.86$ & $3.00$ & $31.30$ & $3.32$ & $0.01$ & $-1.57$ & $0.89$ & OH \\
NGC 2298 & $15.7$ & $-6.30$ & $2.43$ & $20.17$ & $2.88$ & $0.08$ & $-1.85$ & $0.93$ & OH \\
NGC 2419 & $91.5$ & $-9.58$ & $17.88$ & $214.07$ & $2.46$ & $0.03$ & $-2.12$ & $0.86$ & OH \\
Pyxis    & $41.7$ & $-5.75$ & $15.59$ & $71.37$ & $1.04$ & $...$ & $-1.30$ & $-1.00$ & YH \\
NGC 2808 & $11.1$ & $-9.39$ & $2.12$ & $43.42$ & $4.23$ & $0.12$ & $-1.15$ & $-0.49$ & OH \\
E3       & $7.6$ & $-2.77$ & $2.58$ & $13.16$ & $1.41$ & $...$ & $-0.80$ & $...$ & UN \\
Pal. 3   & $95.9$ & $-5.70$ & $17.80$ & $129.70$ & $0.91$ & $...$ & $-1.66$ & $-0.50$ & YH \\
NGC 3201 & $8.9$ & $-7.46$ & $3.90$ & $41.38$ & $2.93$ & $0.12$ & $-1.58$ & $0.08$ & YH \\
Pal. 4   & $111.8$ & $-6.02$ & $17.15$ & $105.78$ & $1.07$ & $...$ & $-1.48$ & $-1.00$ & YH \\
NGC 4147 & $21.3$ & $-6.16$ & $2.41$ & $35.43$ & $2.83$ & $0.08$ & $-1.83$ & $0.66$ & SG \\
NGC 4372 & $7.1$ & $-7.77$ & $6.58$ & $58.75$ & $2.60$ & $0.15$ & $-2.09$ & $1.00$ & OH \\
Rup. 106 & $18.5$ & $-6.35$ & $6.78$ & $31.02$ & $2.01$ & $...$ & $-1.67$ & $-0.82$ & YH \\
NGC 4590 & $10.1$ & $-7.35$ & $4.60$ & $90.02$ & $2.74$ & $0.05$ & $-2.06$ & $0.17$ & YH \\
NGC 4833 & $7.0$ & $-8.16$ & $4.56$ & $33.75$ & $3.08$ & $0.07$ & $-1.80$ & $0.93$ & OH \\
NGC 5024 & $18.3$ & $-8.70$ & $5.75$ & $112.62$ & $3.09$ & $0.01$ & $-1.99$ & $0.81$ & OH \\
NGC 5053 & $16.9$ & $-6.72$ & $16.70$ & $65.21$ & $1.37$ & $0.21$ & $-2.29$ & $0.52$ & YH \\
NGC 5139 & $6.4$ & $-10.29$ & $6.44$ & $87.92$ & $3.63$ & $0.17$ & $-1.62$ & $...$ & UN \\
NGC 5272 & $12.2$ & $-8.93$ & $3.39$ & $115.53$ & $3.64$ & $0.04$ & $-1.57$ & $0.08$ & YH \\
NGC 5286 & $8.4$ & $-8.61$ & $2.21$ & $26.75$ & $3.89$ & $0.12$ & $-1.67$ & $0.80$ & OH \\
AM-4     & $25.5$ & $-1.60$ & $3.65$ & $11.48$ & $0.65$ & $...$ & $-2.00$ & $...$ & UN \\
NGC 5466 & $16.2$ & $-6.96$ & $10.41$ & $158.36$ & $1.88$ & $0.11$ & $-2.22$ & $0.58$ & YH \\
NGC 5634 & $21.2$ & $-7.69$ & $3.96$ & $61.28$ & $3.01$ & $0.02$ & $-1.88$ & $0.91$ & OH \\
NGC 5694 & $29.1$ & $-7.81$ & $3.33$ & $43.30$ & $3.21$ & $0.04$ & $-1.86$ & $1.00$ & OH \\
IC 4499  & $15.7$ & $-7.33$ & $8.25$ & $67.90$ & $2.23$ & $0.08$ & $-1.60$ & $0.11$ & YH \\
NGC 5824 & $25.8$ & $-8.84$ & $3.35$ & $144.28$ & $3.62$ & $0.03$ & $-1.85$ & $0.79$ & OH \\
Pal. 5   & $18.6$ & $-5.17$ & $19.98$ & $109.87$ & $0.60$ & $...$ & $-1.41$ & $-0.40$ & YH \\
NGC 5897 & $7.3$ & $-7.21$ & $7.61$ & $43.54$ & $2.25$ & $0.08$ & $-1.80$ & $0.86$ & OH \\
NGC 5904 & $6.2$ & $-8.81$ & $4.60$ & $61.96$ & $3.33$ & $0.14$ & $-1.27$ & $0.31$ & OH \\
NGC 5927 & $4.5$ & $-7.80$ & $2.54$ & $36.88$ & $3.44$ & $0.04$ & $-0.37$ & $-1.00$ & BD \\
NGC 5946 & $5.8$ & $-7.20$ & $2.13$ & $74.09$ & $3.35$ & $0.16$ & $-1.38$ & $0.69$ & OH \\
BH 176   & $9.7$ & $-4.35$ & $3.92$ & $28.16$ & $1.68$ & $...$ & $0.10$ & $-1.00$ & BD \\
NGC 5986 & $4.8$ & $-8.44$ & $3.18$ & $31.83$ & $3.50$ & $0.06$ & $-1.58$ & $0.97$ & OH \\
Lyng\aa\ 7 & $4.2$ & $...$ & $...$ & $...$ & $...$ & $...$ & $-0.62$ & $-1.00$ & BD \\
Pal. 14  & $69.0$ & $-4.73$ & $24.72$ & $113.07$ & $0.24$ & $...$ & $-1.52$ & $-1.00$ & YH \\
NGC 6093 & $3.8$ & $-8.23$ & $1.89$ & $38.63$ & $3.87$ & $...$ & $-1.75$ & $0.93$ & OH \\
NGC 6121 & $5.9$ & $-7.20$ & $2.34$ & $20.79$ & $3.27$ & $...$ & $-1.20$ & $-0.06$ & OH \\
NGC 6101 & $11.1$ & $-6.91$ & $7.61$ & $32.36$ & $2.13$ & $0.05$ & $-1.82$ & $0.84$ & OH \\
NGC 6144 & $2.6$ & $-6.75$ & $4.01$ & $82.21$ & $2.62$ & $0.25$ & $-1.75$ & $1.00$ & OH \\
NGC 6139 & $3.6$ & $-8.36$ & $2.41$ & $25.03$ & $3.71$ & $0.05$ & $-1.68$ & $0.91$ & OH \\
Terzan 3 & $2.4$ & $-4.61$ & $2.84$ & $12.89$ & $2.07$ & $...$ & $-0.73$ & $-1.00$ & BD \\
NGC 6171 & $3.3$ & $-7.13$ & $5.03$ & $32.47$ & $2.58$ & $0.02$ & $-1.04$ & $-0.73$ & OH \\
1636-283 & $2.0$ & $-3.97$ & $1.30$ & $7.84$ & $2.49$ & $...$ & $-1.50$ & $-0.40$ & YH \\
NGC 6205 & $8.7$ & $-8.70$ & $3.34$ & $56.40$ & $3.56$ & $0.11$ & $-1.54$ & $0.97$ & OH \\
NGC 6229 & $29.7$ & $-8.05$ & $3.27$ & $47.58$ & $3.32$ & $0.05$ & $-1.43$ & $0.24$ & YH \\
NGC 6218 & $4.5$ & $-7.32$ & $3.08$ & $25.09$ & $3.08$ & $0.04$ & $-1.48$ & $0.97$ & OH \\
NGC 6235 & $4.1$ & $-6.44$ & $2.79$ & $25.24$ & $2.81$ & $0.13$ & $-1.40$ & $0.89$ & OH \\
NGC 6254 & $4.6$ & $-7.48$ & $2.32$ & $27.49$ & $3.39$ & $...$ & $-1.52$ & $0.98$ & OH \\
NGC 6256 & $1.8$ & $-6.52$ & $2.08$ & $18.55$ & $3.10$ & $...$ & $-0.70$ & $-1.00$ & BD \\
Pal. 15  & $37.9$ & $-5.49$ & $15.70$ & $64.74$ & $0.93$ & $...$ & $-1.90$ & $1.00$ & OH \\
NGC 6266 & $1.7$ & $-9.19$ & $2.47$ & $18.00$ & $4.02$ & $0.01$ & $-1.29$ & $0.32$ & OH \\
NGC 6273 & $1.6$ & $-9.18$ & $3.13$ & $36.27$ & $3.81$ & $0.27$ & $-1.68$ & $0.96$ & OH \\
NGC 6284 & $7.6$ & $-7.97$ & $3.47$ & $102.72$ & $3.24$ & $0.03$ & $-1.32$ & $0.88$ & OH \\
NGC 6287 & $2.1$ & $-7.36$ & $2.03$ & $28.43$ & $3.46$ & $0.13$ & $-2.05$ & $0.98$ & OH \\
\hline
\label{t:alldata}
\end{tabular}
\end{minipage}
\end{center}
\end{table*}

\setcounter{table}{0}

\begin{table*}
\begin{center}
\begin{minipage}{126mm}
\caption{continued...}
\begin{tabular}{@{}lccccccccc}
\hline \hline
Cluster & $R_{\rm{gc}}$ & $M_V$ & $R_{\rm{h}}$ & $R_{\rm{t}}$ & $\log\,I_{\rm{h}}$ & $\epsilon$ & $[$Fe$/$H$]$ & HB-index & Class \\
Name & (kpc) & & (pc) & (pc) & ($\rm{L}_{\odot}\,\,$pc$^{-2}$) & ($1-b/a$) & & & \\
\hline
NGC 6293 & $1.4$ & $-7.77$ & $2.33$ & $36.43$ & $3.50$ & $0.03$ & $-1.92$ & $0.90$ & OH \\
NGC 6304 & $2.2$ & $-7.32$ & $2.46$ & $23.13$ & $3.28$ & $0.02$ & $-0.59$ & $-1.00$ & BD \\
NGC 6316 & $3.2$ & $-8.35$ & $2.27$ & $18.97$ & $3.76$ & $0.04$ & $-0.55$ & $-1.00$ & BD \\
NGC 6341 & $9.6$ & $-8.20$ & $2.60$ & $36.18$ & $3.58$ & $0.10$ & $-2.28$ & $0.91$ & OH \\
NGC 6325 & $1.1$ & $-6.95$ & $2.19$ & $22.08$ & $3.23$ & $0.12$ & $-1.17$ & $0.84$ & OH \\
NGC 6333 & $1.7$ & $-7.94$ & $2.18$ & $18.75$ & $3.63$ & $0.04$ & $-1.75$ & $0.87$ & OH \\
NGC 6342 & $1.7$ & $-6.44$ & $2.20$ & $37.17$ & $3.02$ & $0.18$ & $-0.65$ & $-1.00$ & BD \\
NGC 6356 & $7.6$ & $-8.52$ & $3.27$ & $35.24$ & $3.51$ & $0.03$ & $-0.50$ & $-1.00$ & BD \\
NGC 6355 & $1.8$ & $-8.08$ & $2.40$ & $41.95$ & $3.60$ & $...$ & $-1.50$ & $0.62$ & OH \\
NGC 6352 & $3.3$ & $-6.48$ & $3.32$ & $17.43$ & $2.68$ & $0.07$ & $-0.70$ & $-1.00$ & BD \\
IC 1257  & $17.9$ & $-6.15$ & $3.17$ & $62.57$ & $2.59$ & $...$ & $-1.70$ & $1.00$ & OH \\
Terzan 2 & $0.9$ & $-5.27$ & $3.85$ & $25.61$ & $2.07$ & $...$ & $-0.40$ & $-1.00$ & BD \\
NGC 6366 & $5.0$ & $-5.77$ & $2.75$ & $15.92$ & $2.56$ & $0.16$ & $-0.82$ & $-0.97$ & OH \\
Terzan 4 & $1.3$ & $-6.09$ & $0.56$ & $4.31$ & $4.07$ & $...$ & $-1.60$ & $1.00$ & OH \\
HP 1     & $6.1$ & $-6.44$ & $6.20$ & $33.71$ & $2.12$ & $...$ & $-1.55$ & $0.75$ & OH \\
NGC 6362 & $5.1$ & $-6.94$ & $4.82$ & $36.85$ & $2.54$ & $0.07$ & $-0.95$ & $-0.58$ & OH \\
Liller 1 & $1.8$ & $-7.63$ & $1.26$ & $35.10$ & $3.98$ & $...$ & $0.22$ & $-1.00$ & BD \\
NGC 6380 & $3.2$ & $-7.46$ & $2.33$ & $37.54$ & $3.38$ & $...$ & $-0.50$ & $-1.00$ & BD \\
Terzan 1 & $2.5$ & $-4.90$ & $6.22$ & $18.03$ & $1.50$ & $...$ & $-1.30$ & $-1.00$ & YH \\
Ton. 2   & $1.4$ & $-6.14$ & $2.54$ & $25.33$ & $2.78$ & $...$ & $-0.50$ & $-1.00$ & BD \\
NGC 6388 & $3.2$ & $-9.42$ & $1.95$ & $18.06$ & $4.32$ & $0.01$ & $-0.60$ & $-1.00$ & BD \\
NGC 6402 & $4.1$ & $-9.12$ & $3.49$ & $89.92$ & $3.69$ & $0.11$ & $-1.39$ & $0.65$ & OH \\
NGC 6401 & $2.7$ & $-7.90$ & $5.83$ & $36.96$ & $2.76$ & $0.15$ & $-0.98$ & $0.35$ & OH \\
NGC 6397 & $6.0$ & $-6.63$ & $1.56$ & $10.58$ & $3.40$ & $0.07$ & $-1.95$ & $0.98$ & OH \\
Pal. 6   & $2.2$ & $-6.81$ & $1.82$ & $14.35$ & $3.33$ & $...$ & $-1.09$ & $-1.00$ & YH \\
NGC 6426 & $14.6$ & $-6.69$ & $5.78$ & $79.66$ & $2.28$ & $0.15$ & $-2.26$ & $0.58$ & YH \\
Djorg. 1 & $4.1$ & $-6.26$ & $4.40$ & $34.98$ & $2.35$ & $...$ & $-2.00$ & $...$ & UN \\
Terzan 5 & $2.4$ & $-7.87$ & $2.49$ & $39.76$ & $3.49$ & $...$ & $0.00$ & $-1.00$ & BD \\
NGC 6440 & $1.3$ & $-8.75$ & $1.42$ & $15.42$ & $4.33$ & $0.01$ & $-0.34$ & $-1.00$ & BD \\
NGC 6441 & $3.9$ & $-9.64$ & $2.18$ & $27.23$ & $4.31$ & $0.02$ & $-0.53$ & $-1.00$ & BD \\
Terzan 6 & $1.6$ & $-7.67$ & $1.22$ & $48.06$ & $4.03$ & $...$ & $-0.50$ & $-1.00$ & BD \\
NGC 6453 & $1.8$ & $-6.88$ & $1.03$ & $60.04$ & $3.86$ & $0.09$ & $-1.53$ & $0.84$ & OH \\
UKS 1    & $0.8$ & $-6.88$ & $2.08$ & $45.29$ & $3.25$ & $...$ & $-0.50$ & $-1.00$ & BD \\
NGC 6496 & $4.3$ & $-7.23$ & $6.26$ & $17.63$ & $2.43$ & $0.16$ & $-0.64$ & $-1.00$ & BD \\
Terzan 9 & $1.6$ & $-3.85$ & $1.47$ & $15.54$ & $2.34$ & $...$ & $-2.00$ & $0.25$ & YH \\
Djorg. 2 & $1.4$ & $-6.98$ & $1.62$ & $20.52$ & $3.50$ & $...$ & $-0.50$ & $-1.00$ & BD \\
NGC 6517 & $4.3$ & $-8.28$ & $1.95$ & $12.88$ & $3.86$ & $0.06$ & $-1.37$ & $0.62$ & OH \\
Terzan 10 & $2.4$ & $-6.31$ & $...$ & $...$ & $...$ & $...$ & $-0.70$ & $-1.00$ & BD \\
NGC 6522 & $0.6$ & $-7.67$ & $2.36$ & $37.30$ & $3.45$ & $0.06$ & $-1.44$ & $0.71$ & OH \\
NGC 6535 & $3.9$ & $-4.75$ & $1.52$ & $16.54$ & $2.67$ & $0.08$ & $-1.80$ & $1.00$ & OH \\
NGC 6528 & $0.6$ & $-6.56$ & $0.99$ & $38.08$ & $3.76$ & $0.11$ & $-0.04$ & $-1.00$ & BD \\
NGC 6539 & $3.1$ & $-8.30$ & $4.08$ & $52.44$ & $3.23$ & $0.08$ & $-0.66$ & $-1.00$ & BD \\
NGC 6540 & $4.4$ & $-5.38$ & $0.26$ & $10.21$ & $4.45$ & $...$ & $-1.20$ & $0.30$ & OH \\
NGC 6544 & $5.3$ & $-6.66$ & $1.39$ & $1.61$ & $3.51$ & $0.22$ & $-1.56$ & $1.00$ & OH \\
NGC 6541 & $2.2$ & $-8.37$ & $2.42$ & $60.27$ & $3.71$ & $0.12$ & $-1.83$ & $1.00$ & OH \\
2MASS-GC01 & $4.5$ & $...$ & $...$ & $...$ & $...$ & $...$ & $-1.20$ & $...$ & UN \\
ESO280-SC06 & $14.3$ & $...$ & $...$ & $...$ & $...$ & $...$ & $-2.00$ & $...$ & UN \\
NGC 6553 & $2.2$ & $-7.77$ & $2.71$ & $14.24$ & $3.37$ & $0.17$ & $-0.21$ & $-1.00$ & BD \\
2MASS-GC02 & $4.1$ & $...$ & $...$ & $...$ & $...$ & $...$ & $...$ & $...$ & UN \\
NGC 6558 & $1.0$ & $-6.46$ & $3.47$ & $22.47$ & $2.63$ & $...$ & $-1.44$ & $0.70$ & OH \\
IC 1276  & $3.7$ & $-6.67$ & $3.69$ & $32.96$ & $2.66$ & $...$ & $-0.73$ & $-1.00$ & BD \\
Terzan 12 & $3.4$ & $-4.14$ & $1.17$ & $4.33$ & $2.65$ & $...$ & $-0.50$ & $-1.00$ & BD \\
NGC 6569 & $2.9$ & $-8.30$ & $4.14$ & $21.63$ & $3.22$ & $...$ & $-0.86$ & $-0.82$ & OH \\
NGC 6584 & $7.0$ & $-7.68$ & $3.12$ & $36.52$ & $3.21$ & $0.03$ & $-1.49$ & $-0.15$ & YH \\
NGC 6624 & $1.2$ & $-7.49$ & $1.88$ & $47.22$ & $3.58$ & $0.06$ & $-0.44$ & $-1.00$ & BD \\
NGC 6626 & $2.7$ & $-8.18$ & $2.54$ & $18.36$ & $3.59$ & $0.16$ & $-1.45$ & $0.90$ & OH \\
NGC 6638 & $2.3$ & $-7.13$ & $1.84$ & $18.51$ & $3.45$ & $0.01$ & $-0.99$ & $-0.30$ & OH \\
NGC 6637 & $1.9$ & $-7.64$ & $2.20$ & $22.10$ & $3.50$ & $0.01$ & $-0.70$ & $-1.00$ & BD \\
NGC 6642 & $1.7$ & $-6.77$ & $1.78$ & $24.61$ & $3.34$ & $0.03$ & $-1.35$ & $-0.04$ & YH \\
NGC 6652 & $2.8$ & $-6.68$ & $1.91$ & $13.16$ & $3.24$ & $0.20$ & $-0.96$ & $-1.00$ & OH \\
NGC 6656 & $4.9$ & $-8.50$ & $3.03$ & $26.97$ & $3.57$ & $0.14$ & $-1.64$ & $0.91$ & OH \\
Pal. 8   & $5.6$ & $-5.52$ & $2.14$ & $50.85$ & $2.68$ & $...$ & $-0.48$ & $-1.00$ & BD \\
\hline
\end{tabular}
\end{minipage}
\end{center}
\end{table*}

\setcounter{table}{0}

\begin{table*}
\begin{center}
\begin{minipage}{126mm}
\caption{continued...}
\begin{tabular}{@{}lccccccccc}
\hline \hline
Cluster & $R_{\rm{gc}}$ & $M_V$ & $R_{\rm{h}}$ & $R_{\rm{t}}$ & $\log\,I_{\rm{h}}$ & $\epsilon$ & $[$Fe$/$H$]$ & HB-index & Class \\
Name & (kpc) & & (pc) & (pc) & ($\rm{L}_{\odot}\,\,$pc$^{-2}$) & ($1-b/a$) & & & \\
\hline
NGC 6681 & $2.1$ & $-7.11$ & $2.43$ & $20.71$ & $3.20$ & $0.01$ & $-1.51$ & $0.96$ & OH \\
NGC 6712 & $3.5$ & $-7.50$ & $2.75$ & $14.93$ & $3.25$ & $0.11$ & $-1.01$ & $-0.62$ & OH \\
NGC 6715 & $19.2$ & $-10.01$ & $3.82$ & $58.23$ & $3.97$ & $0.06$ & $-1.79$ & $0.54$ & SG \\
NGC 6717 & $2.4$ & $-5.66$ & $1.40$ & $20.38$ & $3.10$ & $0.01$ & $-1.29$ & $0.98$ & OH \\
NGC 6723 & $2.6$ & $-7.84$ & $4.07$ & $26.60$ & $3.05$ & $...$ & $-1.12$ & $-0.08$ & OH \\
NGC 6749 & $5.0$ & $-6.70$ & $2.53$ & $11.97$ & $3.00$ & $...$ & $-1.60$ & $1.00$ & OH \\
NGC 6752 & $5.2$ & $-7.73$ & $2.72$ & $64.39$ & $3.35$ & $0.04$ & $-1.56$ & $1.00$ & OH \\
NGC 6760 & $4.8$ & $-7.86$ & $4.69$ & $27.90$ & $2.93$ & $0.04$ & $-0.52$ & $-1.00$ & BD \\
NGC 6779 & $9.7$ & $-7.38$ & $3.41$ & $25.15$ & $3.02$ & $0.03$ & $-1.94$ & $0.98$ & OH \\
Terzan 7 & $16.0$ & $-5.05$ & $6.55$ & $49.06$ & $1.52$ & $...$ & $-0.82$ & $-1.00$ & SG \\
Pal. 10  & $6.4$ & $-5.79$ & $1.70$ & $5.29$ & $2.98$ & $...$ & $-0.10$ & $-1.00$ & BD \\
Arp 2    & $21.4$ & $-5.29$ & $15.89$ & $105.24$ & $0.84$ & $...$ & $-1.84$ & $0.53$ & SG \\
NGC 6809 & $3.9$ & $-7.55$ & $4.46$ & $25.10$ & $2.85$ & $0.02$ & $-1.81$ & $0.87$ & OH \\
Terzan 8 & $19.1$ & $-5.05$ & $7.56$ & $30.25$ & $1.39$ & $...$ & $-1.99$ & $1.00$ & SG \\
Pal. 11  & $7.9$ & $-6.86$ & $5.63$ & $37.13$ & $2.37$ & $...$ & $-0.39$ & $-1.00$ & BD \\
NGC 6838 & $6.7$ & $-5.60$ & $1.92$ & $10.43$ & $2.80$ & $...$ & $-0.73$ & $-1.00$ & BD \\
NGC 6864 & $14.6$ & $-8.55$ & $2.83$ & $43.84$ & $3.65$ & $0.07$ & $-1.16$ & $-0.07$ & OH \\
NGC 6934 & $12.8$ & $-7.46$ & $2.74$ & $38.23$ & $3.24$ & $0.01$ & $-1.54$ & $0.25$ & YH \\
NGC 6981 & $12.9$ & $-7.04$ & $4.35$ & $45.25$ & $2.67$ & $0.02$ & $-1.40$ & $0.14$ & YH \\
NGC 7006 & $38.8$ & $-7.68$ & $4.59$ & $76.54$ & $2.88$ & $0.01$ & $-1.63$ & $-0.28$ & YH \\
NGC 7078 & $10.4$ & $-9.17$ & $3.18$ & $64.42$ & $3.79$ & $0.05$ & $-2.26$ & $0.67$ & YH \\
NGC 7089 & $10.4$ & $-9.02$ & $3.11$ & $71.75$ & $3.75$ & $0.11$ & $-1.62$ & $0.92$ & OH \\
NGC 7099 & $7.1$ & $-7.43$ & $2.68$ & $42.68$ & $3.25$ & $0.01$ & $-2.12$ & $0.89$ & OH \\
Pal. 12  & $15.9$ & $-4.48$ & $7.11$ & $96.78$ & $1.22$ & $...$ & $-0.94$ & $-1.00$ & SG \\
Pal. 13  & $26.7$ & $-3.74$ & $3.45$ & $23.34$ & $1.55$ & $...$ & $-1.74$ & $-0.20$ & YH \\
NGC 7492 & $24.9$ & $-5.77$ & $9.16$ & $62.67$ & $1.51$ & $0.24$ & $-1.51$ & $0.81$ & OH \\
\hline
\end{tabular}
\end{minipage}
\end{center}
\end{table*}

\section{Properties of the different Galactic cluster sub-systems}
\label{s:properties}
It is useful to briefly summarize the results of Mackey \& Gilmore
\shortcite{mackey:04}, who examined some key properties of the Galactic 
globular cluster sub-systems. They found that the majority of globular 
clusters associated with nearby dwarf galaxies are, in terms of HB
morphology and core-radius distribution, essentially indistinguishable
from those in the Galactic YH class. Furthermore, the YH objects are
observed to be characterized by large, energetic orbits around the
Galaxy. These orbits are generally of high eccentricity and intermediate
inclination, and cover a very large range in orbital angular momentum,
including highly retrograde orbits. A small fraction of OH clusters have
similar orbits, but most have much smaller energies and eccentricities.
Many BD clusters (unsurprisingly) have disk-like motions, and/or orbits
which are confined to the Galactic bulge (e.g., Dinescu et al. 
\shortcite{dinescu:03}).   

Mackey \& Gilmore also examined the distribution of cluster ages in
each of the sub-systems, using data from the relative age study of
Salaris \& Weiss \shortcite{salaris:02}. These authors determined relative 
ages for $55$ globular clusters using several different CMD-based 
techniques. A significant fraction of the YH clusters are $\sim 3$ Gyr 
younger than the majority of the OH clusters (see Fig. $9$ 
in Mackey \& Gilmore \shortcite{mackey:04}). However, several OH clusters 
are also seen to be somewhat younger, while a number of YH clusters appear 
to be just as old as the oldest measured OH clusters. The latter result 
is interesting because it hints explicitly at the possible existence of 
a third parameter, in addition to $[$Fe$/$H$]$ and age, which governs 
HB morphology in a cluster.
   
Finally, Mackey \& Gilmore observed that the YH clusters follow a
considerably steeper age-metallicity relationship than do most of the
OH and BD clusters (Mackey \& Gilmore, Fig. $10$). This shows that YH-type
clusters apparently underwent much slower evolution to higher
metallicities than did the OH and BD objects.

Taken together, the results of Mackey \& Gilmore \shortcite{mackey:04} 
fully support the conclusion of Zinn \shortcite{zinn:93} that the YH 
clusters are not native members of the Galaxy, but rather are of external 
origin. Furthermore, Mackey \& Gilmore \shortcite{mackey:04} showed that 
some $15$ per cent of the OH clusters possess 
core radii, orbital motions, ages and HB morphologies which suggest that 
they too are of external origin. The remainder of the OH group, together 
with the BD system, appear to have been formed in the (proto) Galaxy. In 
the following sub-sections, we consider some additional properties of the 
three cluster sub-systems, focusing on structural parameters such as $R_{\rm{h}}$ 
and ellipticity, as well as on cluster luminosities and surface brightnesses.

\subsection{Distribution with Galactocentric radius}
\label{ss:rgcdist}
\begin{figure}
\includegraphics[width=0.5\textwidth]{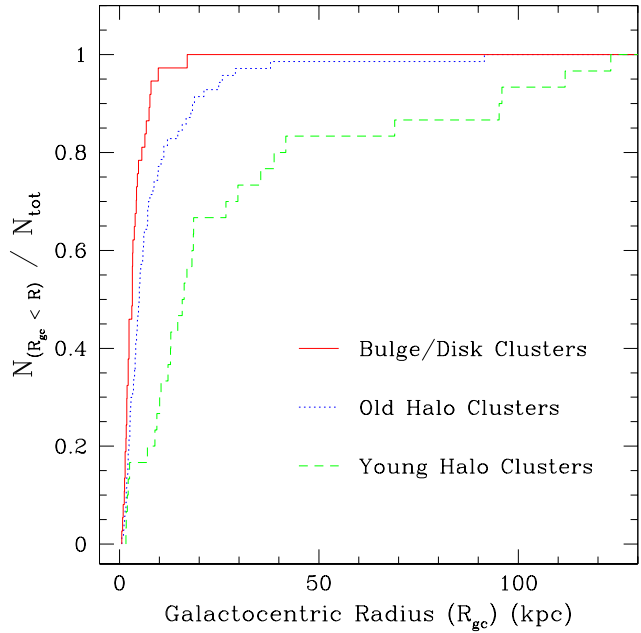}
\caption{Cumulative distributions in Galactocentric radii $R_{\rm{gc}}$ for the
three Galactic globular cluster sub-systems.}
\label{f:cdrgc}
\end{figure}

It has long been known that YH-type clusters typically reside at
much larger Galactocentric distances than do most OH and BD clusters
(see e.g., Zinn \shortcite{zinn:93}; van den Bergh \shortcite{vdb:93}; 
Mackey \& Gilmore \shortcite{mackey:04}). This is clearly seen in 
Fig. \ref{f:cdrgc}, which shows the cumulative distribution in the number 
of globulars as a function of $R_{\rm{gc}}$. Each of the three globular cluster 
sub-systems are quite distinct in terms of their distribution of Galactocentric 
radii. We can quantify the significance of this statement using simple 
Kolmogorov-Smirnov (K-S) tests on each pair of distributions. These show that there 
is only a $2$ per cent chance that the BD and OH samples were drawn from the 
same parent population. The probability that the YH clusters share the
same parent distribution as either the OH or BD clusters is even smaller:
$0.015$ per cent, and $0.01$ per cent, respectively. 

The metal-rich BD clusters are, as expected, the most centrally concentrated 
system, exhibiting a sharp cut-off at $R_{\rm{gc}} = 7.9$ kpc. Only two BD 
clusters lie outside this radius -- BH 176 ($R_{\rm{gc}} = 9.7$ kpc) and 
Pal. 1 ($R_{\rm{gc}} = 17$ kpc). It is interesting to note that both of 
these clusters have been tentatively associated with the candidate dwarf
galaxy in Canis Major, and/or the Monoceros tidal stream (see e.g.,
Frinchaboy et al. \shortcite{frinchaboy:04}). In addition, both 
van den Bergh \& Mackey (2004; hereafter Paper I) and Mackey \& Gilmore 
\shortcite{mackey:04} have previously noted that Pal. 1 has an
unusually high metallicity for its Galactocentric radius. In this
respect it strongly resembles the Sagittarius clusters Ter. 7 and Pal.
12 (see e.g., Fig. $8$ in Mackey \& Gilmore \shortcite{mackey:04}).
Furthermore, examining Table \ref{t:alldata}, Pal. 1 and BH 176 are,
respectively, the lowest luminosity ($M_V = -2.5$) and the third lowest 
luminosity ($M_V = -4.4$) clusters in the BD ensemble. This adds additional
weight to the suspicion that these two objects might not be native BD clusters

The OH clusters are considerably more extended in $R_{\rm{gc}}$ than the BD
objects, and dominate the inner Galactic halo. However, they are not
nearly as extended as the YH clusters, which dominate the outer halo.
Only six of the $70$ OH clusters ($9$ per cent) lie beyond $R_{\rm{gc}} = 20$ 
kpc, and of these, only one (the very unusual cluster NGC 2419 -- 
see Section \ref{ss:cores}) has $R_{\rm{gc}} > 40$ kpc. In contrast, one 
third of the YH system ($10$ clusters) have $R_{\rm{gc}} > 20$ kpc, and six 
of these objects fall outside $R_{\rm{gc}} = 40$ kpc. The most distant YH 
cluster is AM-1 at $R_{\rm{gc}} \sim 120$ kpc. As noted by Mackey \& Gilmore
\shortcite{mackey:04}, the new classifications presented in Table \ref{t:alldata}
show that the YH clusters also extend into even central regions of the Galaxy. 
These objects are therefore not exclusively a remote-halo population.

\subsection{Distribution of cluster luminosities}
\label{ss:mvdist}
\begin{figure}
\includegraphics[width=0.5\textwidth]{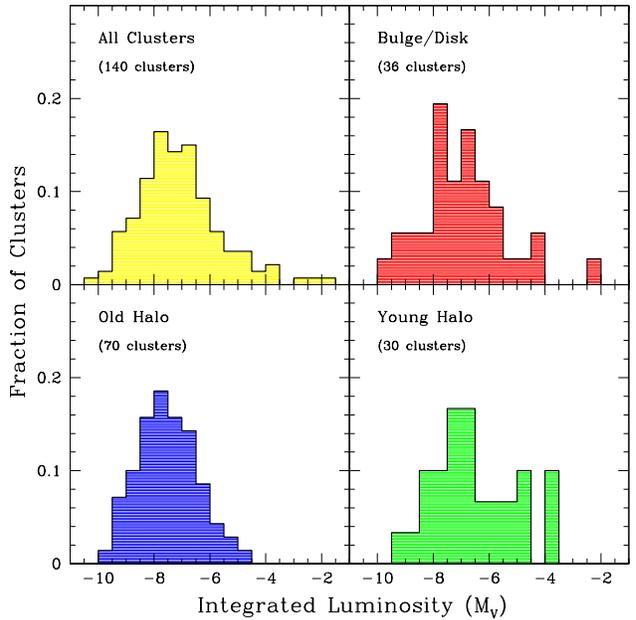}
\caption{Luminosity distributions for the Galactic globular cluster 
sub-systems.}
\label{f:mvdist}
\end{figure}

In Fig. \ref{f:mvdist} we plot histograms of the luminosity distributions 
for clusters in each of the three sub-systems. From this plot it is
seen that both the BD and OH systems have peaks near $M_V \sim -7.75$.
Furthermore, these two systems contain very few low luminosity clusters.
Only $11$ per cent of BD clusters and $4$ per cent of OH clusters are 
fainter than $M_V = -5.5$. This contrasts with the situation for the YH 
system, which contains a significant population of low luminosity objects. 
It is seen that $27$ per cent of the clusters in this sub-system are fainter 
than $M_V = -5.5$. In addition, the distribution of YH clusters peaks at somewhat 
lower luminosity ($M_V \sim -7$) than do the OH and BD distributions.
Again, we can evaluate the significance of the differences between the three
distributions using K-S tests. These show that there is only a $5$ per cent
chance that the OH and YH clusters share the same parent luminosity distribution.
However, the K-S tests are unable to discriminate between the OH and BD distributions,
and the YH and BD distributions -- the results do not confirm statistically 
significant differences between these parent populations, but they also do not
confirm statistically significant similarities. This result is at least partly due
to the relatively small BD and YH sample sizes. As discussed above, the primary 
difference between the distributions is the presence (or not) of a small percentage
of low luminosity clusters. Hence, it would be necessary for the sample sizes to be
larger to confirm (or deny) this difference at a significant level. This is, of course, 
not possible, since we included all known Galactic globular clusters in Table 
\ref{t:alldata}.

We can, however, clearly state that the luminosity distribution of YH clusters is
significantly broader than that for the OH clusters. Combined with our observations
from Section \ref{ss:rgcdist}, this result is similar to that obtained by
Kavelaars \& Hanes \shortcite{kavelaars:97}, who demonstrated that outer halo
globular clusters (i.e., those metal-poor clusters with $R_{\rm{gc}} > 8$ kpc)
have a broader luminosity function than the inner halo
globular clusters (i.e., those metal-poor clusters with $R_{\rm{gc}} < 8$ kpc).
The present result is, however, distinct from this, as we have not used $R_{\rm{gc}}$
as a discriminator -- indeed, Fig. \ref{f:cdrgc} shows that there is a significant
population of OH clusters with $R_{\rm{gc}} > 8$ kpc, and vice versa for YH clusters.
Our observations imply that LF broadness is intrinsic to cluster sub-population
rather than simply location within the Galaxy.

As discussed by Kavelaars \& Hanes \shortcite{kavelaars:97}, it seems likely 
that this observed difference in broadness, as well as the apparent dearth of low luminosity 
clusters in the OH and BD systems, can (at least in part) be attributed to the 
destructive influence of tidal forces and bulge and disk shocks in the inner Galaxy. 
A number of authors (see e.g., Fall \& Zhang \shortcite{fall:01}) have shown that 
gravitational shocks, in addition to cluster relaxation and evaporation, 
are important in shaping the globular cluster mass function. Since 
the YH clusters typically lie at larger Galactocentric radii than do OH and BD 
clusters, they should be much less strongly affected by destructive external forces. 

In addition, if the accretion hypothesis is correct, many of the YH clusters 
have spent some fraction of their life outside the Galaxy in a more benign tidal 
environment. Consider, for example, the clusters associated with the 
Sagittarius dwarf. While it is at present very difficult to constrain the 
early orbital evolution of this galaxy, models by Jiang \& Binney 
\shortcite{jiang:00} have shown that its Galactocentric distance at 
apocentre $\sim 11$ Gyr ago most likely fell in the range 
$60 < R_{\rm{gc}} < 250$ kpc, with the exact value depending on its initial 
mass (specifically, the amount of dark matter it contained). Models by 
Helmi \& White \shortcite{helmi:01} and Ibata et al. \shortcite{ibata:01},
which assume comparatively small initial masses, show that Sagittarius may
have followed a similar orbit to that presently observed (which has
apocentric radius $\sim 65$ kpc, pericentric radius $\sim 15$ kpc, and 
an orbital period $\sim 1$ Gyr) for the last $\sim 10-12$ Gyr. In each 
of these scenarios, the Sagittarius dwarf, and its remaining globular 
clusters, spend a significant fraction of their lives away from the inner 
regions of the Galaxy. The observation that four of the Sagittarius
clusters are still associated with the main body of this galaxy provides 
further evidence that these clusters have not suffered prolonged tidal 
stresses due to the Galaxy. Tidal forces strong enough to disrupt a 
globular cluster are likely to be so strong that they would also pull 
that cluster out of its parent dwarf galaxy.

Together the factors descibed above may help to explain 
why the YH system has a greater fraction of low luminosity clusters than 
do the OH and BD systems. If destructive effects are indeed responsible,
then the YH luminosity distribution may well indicate that globular clusters
are formed with a broad LF. Alternatively, as suggested by Kavelaars \& Hanes
\shortcite{kavelaars:97}, the shape of the cluster LF may be dependent on
the environment in which clusters form.
   
\begin{figure}
\includegraphics[width=0.5\textwidth]{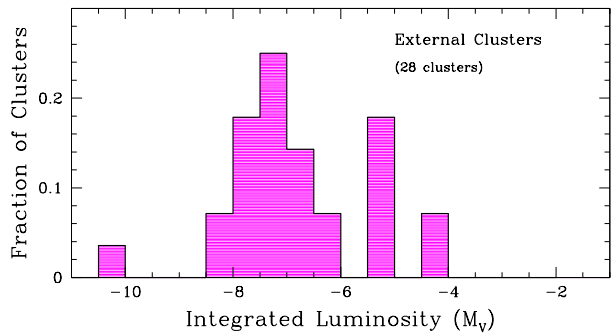}
\caption{Combined luminosity distribution for the globular clusters
in the LMC, SMC and Fornax and Sagittarius dwarf galaxies.
Note how similar this distribution is to that for the YH clusters
(Fig. \ref{f:mvdist}).}
\label{f:mvdistex}
\end{figure}

Finally, it is interesting to examine the combined luminosity
distribution of globular clusters in the four nearby dwarf galaxies
studied by Mackey \& Gilmore \shortcite{mackey:04}. This distribution 
is plotted in Fig. \ref{f:mvdistex}, and bears a strong resemblance 
to the luminosity distribution of the YH clusters. A significant 
fraction of the external globular clusters ($25$ per cent) are fainter than 
$M_V = -5.5$, and in fact this fraction is almost identical to that observed 
for the YH clusters. A K-S test gives a $92$ per cent chance that the
YH and external clusters have the same parent luminosity distribution,
but only a $3$ per cent chance that the OH and external clusters share
the same parent luminosity distribution.
This observation adds another example to the list of
characteristics in which the YH sub-system closely resembles the group 
of globular clusters associated with nearby Local Group dwarf
galaxies \cite{mackey:04}. It therefore offers further evidence in favour of the
hypothesis that much, or all, of the ensemble of YH clusters is of external origin.

\subsection{Distribution of half-light and tidal radii}
\label{ss:rhdist}
\begin{figure}
\includegraphics[width=0.5\textwidth]{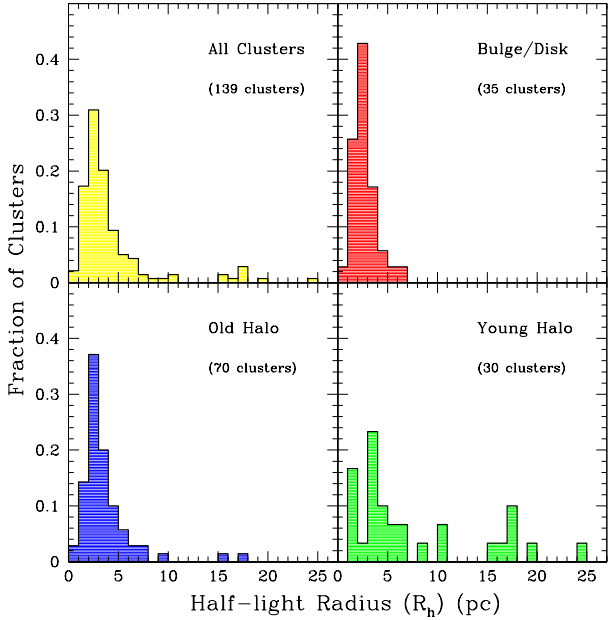}
\caption{Distributions of half-light radii ($R_{\rm{h}}$) for the Galactic 
globular cluster sub-systems.}
\label{f:rhdist}
\end{figure}

\begin{figure}
\includegraphics[width=0.5\textwidth]{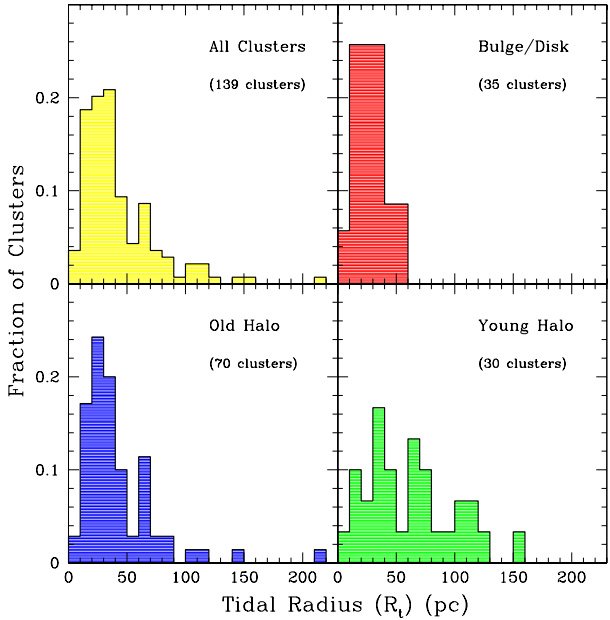}
\caption{Distributions of tidal radii ($R_{\rm{t}}$) for the Galactic
globular cluster sub-systems.}
\label{f:rtdist}
\end{figure}

Figs. \ref{f:rhdist} and \ref{f:rtdist} show the distributions of 
half-light radii ($R_{\rm{h}}$) and tidal radii ($R_{\rm{t}}$), respectively, for the 
three Galactic cluster sub-systems. Just as Mackey \& Gilmore 
\shortcite{mackey:04} demonstrated for core
radii, it is clear from these two Figures that the majority of
diffuse clusters -- that is, clusters with large half-light and
tidal radii -- fall into the YH category. Conversely, the large
majority of OH and BD objects are seen to be compact. Once again, we
apply K-S tests to determine the significance of the differences between
the three distributions. These calculations show that there is a less than
$1$ per cent chance that the OH and YH distributions in $R_{\rm{h}}$ were drawn 
from the same parent population, and a less than $0.2$ per cent chance that the 
BD and YH distributions in $R_{\rm{h}}$ share the same parent population. 
For the BD and OH $R_{\rm{h}}$ distributions, this probability is $8$ per cent. 
Exactly similar results are obtained for the $R_{\rm{t}}$ distributions.

Given what is known about the distributions in Galactocentric radius for the
three sub-systems (Section \ref{ss:rgcdist}), this result is not unexpected when
considering $R_{\rm{t}}$.
The tidal radii of clusters are determined by how deeply their orbits
dive into the the Galactic potential in which they reside (see e.g.,
King \shortcite{king:62}). In fact, it has been shown that $R_{\rm{t}}$ is most 
strongly correlated with the distance of a cluster at the pericentre 
of its orbit \cite{vdb:94}. Hence, since the majority of clusters
which reside at large $R_{\rm{gc}}$ belong to the YH ensemble, it is to
be expected that this group also contains the largest fraction of
clusters with large $R_{\rm{t}}$. By the same token, the BD clusters are the
most centrally concentrated in $R_{\rm{gc}}$, and consequently all the
clusters in this sub-sample have comparatively small tidal radii.

The distributions in $R_{\rm{h}}$ are somewhat more intriguing. It has long
been known that the half-light radii of Galactic globular clusters
correlate with their Galactocentric distances, in the sense that the
most remote clusters are also the largest (e.g., van den Bergh \& Morbey
\shortcite{vdb:84}). Hence, with only this fact in mind, one might expect 
the YH system to possess most of the clusters with large $R_{\rm{h}}$. This
is indeed what is observed. However, the fact that the YH clusters are
strongly suspected of being accreted in a number of minor merger events
presents a challenge for this explanation. This is so because, for
clusters which have been more-or-less randomly accreted, there is no
reason to expect half-light radii to follow the same trend with $R_{\rm{gc}}$
as that which is observed for clusters which are native to the Galaxy.
We address this puzzle in more detail in Section \ref{ss:comphalo}.

Irrespective of the above result, the distributions in $R_{\rm{h}}$ are
interesting because, as described in Section \ref{s:intro}, a cluster's 
half-light radius is stable over many relaxation times. Hence, the 
observed distribution of half-light radii for the YH clusters should, to
a large extent, reflect the distribution of radii with which these
clusters were formed. If this is the case, then the fact that the OH
and BD systems contain almost no diffuse clusters with $R_{\rm{h}} > 7$ pc 
allows two possibilities: either (i) the distributions in $R_{\rm{h}}$ for these
two ensembles also remain mostly unmodified from their original states,
and the dearth of diffuse objects is due to environmental influence on
cluster formation (i.e., no diffuse clusters were formed); or (ii) the native 
Galactic cluster population has been strongly modified by destructive tidal forces.
From the cluster destruction studies of Gnedin \& Ostriker \shortcite{gnedin:97}
and Dinescu, Girard, \& van Altena \shortcite{dinescu:99} it is clear that
destructive influences, such as evaporation due to two-body relaxation as well
as disk and bulge shocks, must be at least partly responsible. These 
most strongly affect diffuse clusters -- that is, objects with large $R_{\rm{h}}$
and low $M_V$. For example, in the formalism of Dinescu et al. \shortcite{dinescu:99}
the destruction rate due to both bulge and disk shocks is directly proportional
to $R_{\rm{h}}^3$, and inversely proportional to cluster mass. Furthermore,
clusters with short orbital periods and small pericentric radii (i.e., many OH
clusters and most BD clusters) are most strongly affected. In their study,
Dinescu et al. estimated destruction rates for two of our BD clusters, $25$ of
our OH clusters, and $10$ of our YH clusters. They provide a list of the nine
clusters with the largest destruction rates (these objects have expected survival
times of $\sim 10$ Gyr or less). Of these, one is a BD cluster, $7$ are OH 
clusters, and one a YH cluster. Clearly the native population is most strongly
affected. In addition, several of the nine have large $R_{\rm{h}}$ (e.g., 
NGC 288, Pal. 5, and NGC 6144), while two (NGC 6121 and 6397) have been
shown to have depleted luminosity functions at low stellar masses
\cite{kanatas:95,piotto:97}. This is expected in a cluster if it has been subjected
to strong external tidal forces.

Assuming that some fraction of the native Galactic clusters has been destroyed
as discussed above, it is possible to place constraints on the initial population.
Examining Fig. \ref{f:rhdist}, one sees that $33$ per cent of YH clusters 
have $R_{\rm{h}} > 7$ pc. Mackey \& Gilmore \shortcite{mackey:04} concluded that 
there are presently $\sim 100$ native Galactic members in the BD/OH set 
(recall that $\sim 15$ per cent of the OH clusters are possibly of external 
origin). If the initial distribution of cluster sizes for native Galactic 
globulars resembled that presently observed for YH objects, then this 
population constitutes only $67$ per cent of the original population, since 
essentially no OH or BD clusters are observed to have $R_{\rm{h}} > 7$ pc. Hence 
the original population of native Galactic globular clusters may have 
numbered $\sim 150$ objects. This result is consistent with both the 
estimate of Mackey \& Gilmore \shortcite{mackey:04} who postulated an original
Galactic population numbering $\sim 200$ objects, and the result of Gnedin \&
Ostriker \shortcite{gnedin:97} who concluded that up to half of the present 
population of globular clusters are likely to be destroyed during the next 
Hubble time.
   
It is important to point out that the above argument explicitly
depends on the assumption that the OH and BD systems originally had
$R_{\rm{h}}$ distributions similar to that which is presently observed for YH
clusters. Alternatively we again note the possibility that formation
conditions did not allow the production of diffuse clusters in the
inner Galaxy. In that case our estimate of $\sim 150$ original Galactic
globular clusters is an over-estimate. However, not enough information
about ancient cluster formation conditions is available at present to
allow this possibility to be confirmed or rejected.

\subsection{Distribution of surface brightnesses}
\label{ss:sbdist}
\begin{figure}
\includegraphics[width=0.5\textwidth]{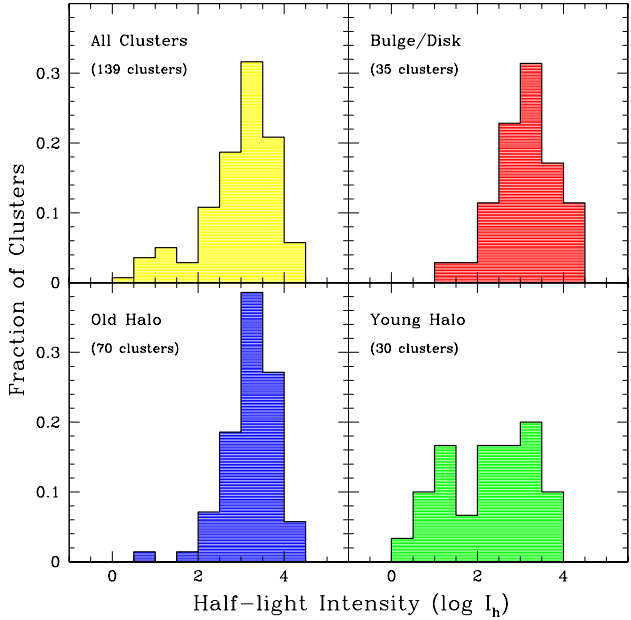}
\caption{Distributions of half-light intensity ($I_{\rm{h}}$) for
the three Galactic cluster sub-systems.}
\label{f:sbdist}
\end{figure}

In Fig. \ref{f:sbdist} we plot the distributions of $I_{\rm{h}}$, the half-light
intensity (essentially equivalent to the mean surface brightness within $R_{\rm{h}}$), 
for the three Galactic globular cluster sub-systems. Since this quantity is closely 
related to both $R_{\rm{h}}$ and $M_V$, it is not surprising to see that the vast 
majority of low surface brightness clusters belong to the YH sub-system.
The distribution of $I_{\rm{h}}$ for this ensemble is quite broad with no
prominent maximum. In this respect it differs significantly from the much sharper 
singly-peaked distributions that are observed for the BD and OH clusters. A 
Kolmogorov-Smirnov test shows that there is only a $1$ per cent chance that the 
YH and BD samples were drawn from the same parent distribution, and a
less than $0.1$ per cent probability that the YH and OH samples were drawn
from the same parent distribution.

The lack of low surface brightness clusters among the OH and BD
systems results in a scarcity of clusters with $I_{\rm{h}} < 100$ $L\odot\,\,$pc$^{-2}$
and $R_{\rm{gc}} < 15$ kpc (as can be seen by inspecting Table \ref{t:alldata}). 
It is also of interest to note that all the nine known Galactic globular clusters 
with $I_{\rm{h}} < 20$ $L\odot\,\,$pc$^{-2}$ are metal-poor objects having 
$[$Fe$/$H$] < -1.2$. As was already pointed out in Sections \ref{ss:mvdist} 
and \ref{ss:rhdist}, these effects are likely due to the tidal destruction 
of fragile low surface brightness clusters that ventured too close to the 
massive bulge of the Galaxy. An additional complication is that observational 
selection effects militate against the discovery of low surface brightness 
clusters against the rich foreground of stars in the direction of the Galactic
bulge.
   
It is interesting to note that the location of the peak surface
brightness for the OH and BD clusters coincides closely with the
location of the (marginal) surface brightness peak for the YH clusters,
at $\log\,I_{\rm{h}} \sim 3.25$. As previously noted, the YH population
has apparently not been as strongly affected by tidal forces as have
the OH and BD populations. This sub-system may therefore provide
interesting constraints on the initial distribution of cluster surface
brightnesses (although, as noted previously, $I_{\rm{h}}$ is more sensitive
to cluster evolution than is $R_{\rm{h}}$, because the evaporation of
stars, along with stellar evolution, causes variations in $M_V$). Thirty-seven
per cent of YH clusters have $I_{\rm{h}} < 100$ $L\odot\,\,$pc$^{-2}$, while for
the BD and OH clusters this fraction is essentially zero. If we again
set the number of native Galactic members in the BD and OH groups to be
$\sim 100$ then, assuming the distribution of surface brightnesses for this
ensemble used to resemble that now observed for the YH clusters, the
present population of native Galactic clusters represents $63$ per cent of 
the original population, which must therefore have numbered $\sim 160$ 
clusters. This is consistent with the estimate already given in the previous
Section. It is interesting to note that in a thought experiment where
we assume all YH clusters with $I_{\rm{h}} < 100$ mag pc$^{-2}$ to have been 
destroyed, the amplitude of the surface brightness peak at $\log\,I_{\rm{h}} \sim 3.25$
rises to $\sim 0.32$ -- which is similar to the amplitudes of the peaks 
that are observed for the OH and BD clusters.

\subsection{Cluster ellipticities}
\label{ss:ellip}
\begin{figure}
\includegraphics[width=0.5\textwidth]{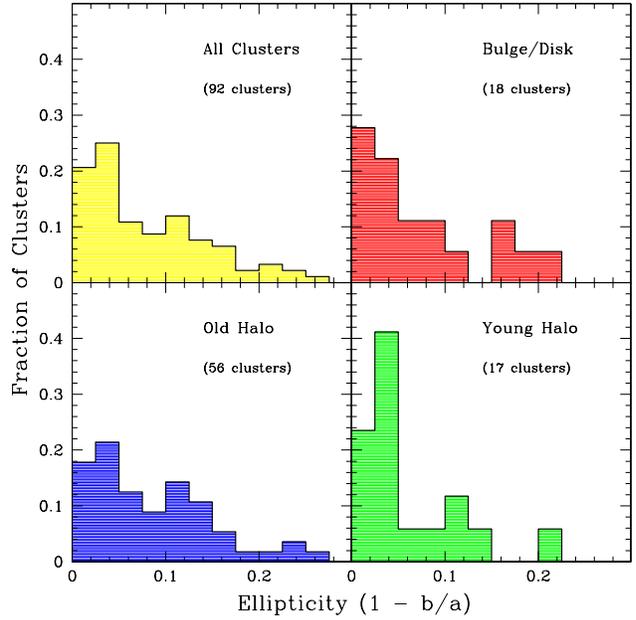}
\caption{Distributions of ellipticity for the three Galactic globular 
cluster sub-systems.}
\label{f:ellip}
\end{figure}

Next we consider the ellipticity distributions of Galactic globular 
clusters, which are plotted in Fig. \ref{f:ellip}. This Figure shows some 
interesting features. In particular, the YH clusters appear on average to 
be somewhat less flattened than the OH clusters, although the small YH
sample precludes this difference being confirmed as significant by
a K-S test. Additional measurements will be required to further investigate 
this possible difference. If the average flattening of YH clusters does turn out
to be less than that of OH clusters, this would be unexpected for several reasons. 
Frenk \& Fall \shortcite{frenk:82} found the 
oldest LMC globular clusters to be slightly {\em more} elliptical than typical 
Galactic globular clusters. A similar result was obtained by Goodwin 
\shortcite{goodwin:97}. Since a number of other parameters show a strong 
similarity between the YH clusters and the sample of external clusters 
(of which LMC objects make up a significant fraction) it is perhaps 
surprising that the YH clusters should be less elliptical on average 
than the OH clusters. Frenk \& Fall \shortcite{frenk:82} also found 
that older LMC clusters are typically less elliptical than younger LMC 
clusters. If this were true in general, one would have expected the YH clusters, 
many of which are $\sim 3$ Gyr younger than most of the OH clusters, to be 
more elliptical than these Galactic natives.
   
Goodwin \shortcite{goodwin:97} has argued that the strength of the tidal 
field in which a cluster evolves is the predominant factor in shaping its
ellipticity. This is so because a strong tidal field is quite effective
in reducing the triaxiality of a cluster and in removing any angular
momentum it may have, thus reducing its ellipticity. This may explain
why clusters in the LMC are more elliptical, on average, than Galactic
clusters. However, this prediction appears to be at odds with the
present observation that the YH clusters, which generally reside at
large $R_{\rm{gc}}$ and have supposedly spent at least some fraction of their
lifetimes in low-mass galaxies, may typically be less flattened than are
the OH and BD clusters, which reside in a strong tidal field. One
possible explanation for this puzzle is that recent bulge and disk
shocks have re-introduced velocity anisotropy into many of the OH and
BD clusters \cite{goodwin:97}. If so, this might explain the tail to high
ellipticities for the OH clusters, which is not present for the
YH clusters.

Finally, it is interesting to note that all three of the lowest
surface brightness ($I_{\rm{h}} < 30$ $L_\odot\,$pc$^{-2}$) clusters for which
ellipticity values have been published are unusually flattened objects
with $1 - b/a > 0.2$. These clusters are NGC 5053, NGC 7492, and Pal. 1.
Examination of Digitized Sky Survey (DSS) images confirms this unusually
high ellipticity for NGC 5053 and 7492 (the DSS image of Pal. 1 is too
faint for useful measurements), so this result does not seem to be a
measurement artifact due to low surface brightness. Clearly, more
observational data would be valuable. Such information could be
obtained in the near future because there are still a significant number
of low surface brightness clusters for which ellipticity measurements
are not yet available. In this connection, it is of interest to note that
the very faint ($M_V = -3.0^{+2.0}_{-0.7}$) and large ($R_{\rm{h}} = 23 \pm 10$ pc) 
globular cluster-like object recently discovered in the outer halo of the 
Galaxy by Willman et al. \shortcite{willman:04} is of exceedingly low 
half-light intensity ($I_{\rm{h}} = 0.40^{+2.00}_{-0.37}$ $L_\odot\,$pc$^{-2}$ 
-- i.e., $\log I_{\rm{h}} = -0.4^{+0.8}_{-1.1}$), and also 
appears to be significantly flattened in its outer regions.
   
If the observed ellipticity of the lowest surface brightness
globulars were due to rotation, then this would hint at a physical
relation between rotation and surface brightness. Alternatively, it is
possible that some globular clusters might be embedded in small dark
matter halos so that the shapes of the lowest surface brightness
clusters reflect the axis ratios of their dark matter mini-halos. In this
connection it is noted that Mashchenko \& Sills 
\shortcite{mashchenko:04a,mashchenko:04b} have recently
investigated the formation and evolution of globular clusters in
different types of dark matter mini-halos. They conclude that the
presence of obvious tidal tails is the only observational evidence that 
can reliably rule out significant amounts of dark matter in a globular
cluster. Furthermore, they write that {\em ``cosmological dark matter halos
are known to have noticeably non-spherical shapes; a stellar cluster
relaxing inside such a halo would have a spherical distribution in its
denser part where stars dominate dark matter, and would exhibit
isodensity contours of increasingly larger eccentricity in its
outskirts where DM becomes the dominant mass component.''} It is currently
believed (e.g., Bailin \& Steinmetz \shortcite{bailin:04}) that
massive cosmological dark halos are triaxial with $c/a$ ratios of $\sim 0.6$.
However, the numerical resolution on cosmological $N$-body simulations
is presently insufficient to reliably predict the likely flattenings
of subhalos with masses that are only a millionth of those of the
massive halos themselves. Furthermore, the shapes of such dark matter
subhalos orbiting the Galaxy may have been strongly affected by tidal
stripping, which rapidly reduces the masses of dark matter subhalos
once they have been accreted onto larger structures (e.g., Gao et al.
\shortcite{gao:04}). Even so, the case for a detailed new systematic study 
of globular cluster ellipticities is strong, especially because of the 
present availability of sensitive wide field cameras at a number of large 
telescopes. Such a study should certainly include low surface brightness 
clusters, because such objects are still very poorly studied. It would be
particularly interesting to examine how ellipticity changes with
projected radius, i.e. do clusters get more or less flattened with
increasing projected radius?

\section{The log $\mathbf{\rm{R}_{\rm{h}}}$ versus $\mathbf{\rm{M_V}}$ plane}
\label{s:rhmvplane}

\subsection{The link between cluster size and luminosity}
\label{ss:mvrhlink}

\begin{figure}
\includegraphics[width=0.5\textwidth]{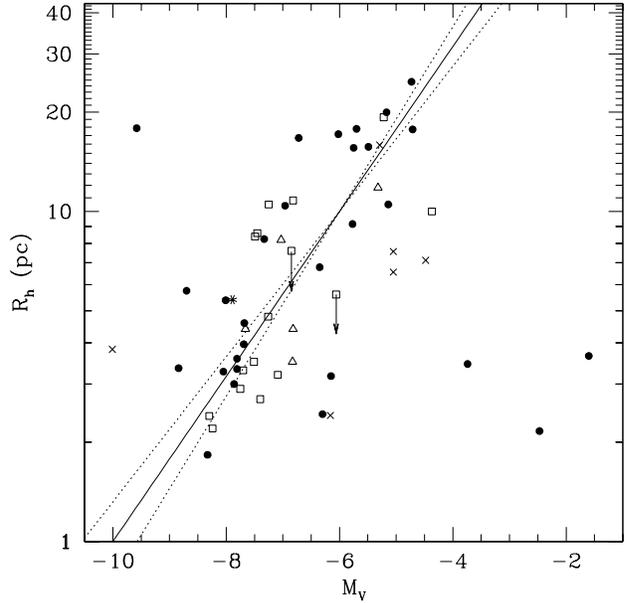}
\caption{Plot of $\log R_{\rm{h}}$ vs. $M_V$ for Galactic globular clusters
with $R_{\rm{gc}} > 15$ kpc (filled circles), and those in the LMC (open
squares), SMC (star), and Fornax and Sagittarius dwarf galaxies (open
triangles, and crosses, respectively). The best fit line, Eq. 
\ref{e:rhmv} is marked by the solid line, while the two $1\sigma$ error
lines are marked by the dotted lines.}
\label{f:rhmvouter}
\end{figure}

It is informative to consider the correlation between cluster
luminosity and half-light radius. Several authors have examined this
relationship in the past, arriving at different conclusions. For example, 
Ostriker \& Gnedin \shortcite{ostriker:97} found $R_{\rm{h}}$ to be strongly 
correlated with luminosity\footnote{assuming a roughly constant $M/L$ ratio for 
Galactic globular clusters} for clusters with $5 < R_{\rm{gc}} < 60$ kpc, such 
that $R_{\rm{h}} \propto (L/L_{\odot})^{-0.63}$ with correlation coefficient 
$0.77$. However, van den Bergh, Morbey \& Pazder \shortcite{vdb:91} found
that half-light radius also correlates quite strongly with
Galactocentric radius $R_{\rm{gc}}$ for clusters with $R_{\rm{gc}} < 40$ kpc: 
$R_{\rm{h}} \propto R_{\rm{gc}}^{0.5}$ with correlation coefficient 
$0.37^{+0.16}_{-0.19}$. They found that if this dependence is taken into 
account, cluster size appears uncorrelated with luminosity -- that is, a 
``normalized'' radius $R_{\rm{h}}^*$, which is the size a cluster would have at 
given $R_{\rm{gc}}$, does not correlate with luminosity. A similar conclusion 
was reached by McLaughlin \shortcite{mclaughlin:00} who found that for the 
full sample of Galactic globular clusters, a weak dependence of $R_{\rm{h}}$ 
on luminosity disappears when the variation of $R_{\rm{h}}$ on $R_{\rm{gc}}$ 
(McLaughlin found $R_{\rm{h}} \propto R_{\rm{gc}}^{0.4}$) is taken into account.
However, McLaughlin stresses that he cannot reproduce the result of
Ostriker \& Gnedin \shortcite{ostriker:97} for clusters with 
$5 < R_{\rm{gc}} < 60$ kpc.

It seems possible that these different conclusions are due, at least in part,
to different sample selections. McLaughlin \shortcite{mclaughlin:00} considered
the full Galactic globular cluster system, and the sample of van den Bergh et al.
\shortcite{vdb:91} contained $98$ clusters with $R_{\rm{gc}} < 40$ kpc.
In contrast, the ensemble of Ostriker \& Gnedin \shortcite{ostriker:97} consisted
of $58$ non-core-collapsed clusters with $5 < R_{\rm{gc}} < 60$ kpc.
In Section \ref{s:properties}, we discussed the idea that the globular cluster
sample consisting of objects in the central Galaxy has had its properties 
strongly modified by external (tidal) influences. We found support for this notion,
in agreement with a number of previous authors. With this in mind, in the
present context it makes sense to consider a cluster sample which is as free
from tidal modification as possible. To this end, we consider a Galactic globular
cluster sample consisting only of objects with $R_{\rm{gc}} > 15$ kpc. 
The majority of clusters in this ensemble are YH members.
In Paper I we made plots of $\log R_{\rm{h}}$ versus $M_V$ for both this Galactic 
sample, as well as for a sample consisting of globular clusters in the LMC, SMC, 
and Fornax and Sagittarius dwarf galaxies (Figs. 6 \& 7 of Paper I). Interestingly, 
both of these plots appear to show a strong inverse correlation between cluster size and 
luminosity, particularly for clusters in the external dwarf galaxies. We reproduce 
these data here, in Fig. \ref{f:rhmvouter}.
   
Apart from four notable outliers, the majority of clusters in
Fig. \ref{f:rhmvouter} exhibit a clear trend in $R_{\rm{h}}$ with $M_V$. Three 
of the deviant clusters -- Pal. 1, AM-4, and Pal. 13 -- are compact but 
of very low luminosity, and lie below and to the right of the majority of 
the data. All three of these objects have been shown to be extremely poorly
populated globular clusters -- AM-4 does not even possess any RGB or HB
stars \cite{inman:87}. Pal. 1 has been studied by Rosenberg et
al. \shortcite{rosenberg:98}, who measured an unusually flat mass function, 
very different from those measured for $21$ more luminous Galactic globular 
clusters. These authors concluded that Pal. 1 has most likely been modified
by tidal forces in combination with the evaporation of cluster members. Assuming 
this to be the case, they find Pal. 1 must presently be very close to its final 
destruction. Similarly, Siegel et al. \shortcite{siegel:01} showed that Pal. 13 
has been strongly affected by tidal forces and is on the threshold of dissolution, 
a result supported by the unexpectedly high internal velocity dispersion measured 
for this object by C\^ot\'e  et al. \shortcite{cote:02}. We note that the inner 
halo cluster E3 also inhabits this region of the $\log R_{\rm{h}}$ vs. $M_V$ plane. 
This cluster is also exceedingly sparse. The photometric study by McClure et al. 
\shortcite{mcclure:85} found a sharp drop-off in the cluster luminosity 
function several magnitudes below the main sequence turn off. This supports 
the suggestion by van den Bergh et al. \shortcite{vdb:80} that E3 has been 
strongly modified by evaporative processes.
   
Since Pal. 1, AM-4 and Pal. 13 have apparently been strongly affected by external 
destructive forces, we exclude them from our present sample and analysis. All three 
clusters are apparently very close to dissolution, so the lower right hand 
corner of Fig. \ref{f:rhmvouter} may constitute the ultimate ``graveyard'' of compact 
globular clusters. This leaves the anomalous object NGC 2419, which is the fourth 
deviant cluster in Fig. \ref{f:rhmvouter}, and which is discussed in more detail in 
Section \ref{ss:cores}, below. We note that the inclusion or omission of this object 
from the sample does not strongly affect our results (below), although the formal 
errors are reduced if NGC 2419 is not included. 

Applying a least-squares analysis to the sample, one obtains a best-fit line:
\begin{equation}
\log R_{\rm{h}} = 0.25 M_V + 2.5,
\label{e:rhmv}
\end{equation}
where the formal errors are $\pm 0.03$ on the slope and $\pm 0.18$ on the
intercept. This line is plotted in Fig. \ref{f:rhmvouter}, along with the 
$1\sigma$ error lines. These errors represent the case when NGC 2419 is not
included in the analysis. The trend defined by Eq. \ref{e:rhmv} is strong, 
with correlation coefficient $0.83$. It is important to note however, that
the distribution of cluster radii in Fig. \ref{f:rhmvouter} is dominated by scatter.

If one replaces the integrated magnitude in Eq. \ref{e:rhmv} with 
$M_V = -2.5\log(L/L_{\odot}) + M_{V,\odot}$ one finds that 
$R_{\rm{h}} \propto (L/L_{\odot})^{-0.625}$, which is almost exactly 
the result of Ostriker \& Gnedin \shortcite{ostriker:97} if $M/L$ is 
assumed to be approximately constant for Galactic globular clusters.
The physical significance of Eq. \ref{e:rhmv} is not clear, however.
One possibility is that the relationship between $R_{\rm{h}}$ and $M_V$ 
has been imposed purely through the dynamical evolution of clusters. If this is
the case, then the majority of evolution must have been in $M_V$ as
$R_{\rm{h}}$ is stable over cluster lifetimes (see Section \ref{s:intro}).
$M_V$ fades as a cluster evolves, due to both stellar evolution and
the evaporation of cluster members. The rate of fading due to stellar evolution
is regulated by the cluster IMF (because more massive stars die earlier),
while evaporation is caused by both external influences (e.g., tidal stripping)
and internal relaxation. Loosely bound clusters are susceptible
to both processes (see e.g., Gnedin \& Ostriker \shortcite{gnedin:97}),
although we note that our sample has been chosen to minimise possible
modifications due to tidal shocking. In contrast, compact clusters are 
significantly more stable, although very compact objects suffer accelerated 
evaporation due to internal relaxation (see e.g., Figs. 6 \& 7 in 
Gnedin \& Ostriker \shortcite{gnedin:97}). Given the complexity of these
combined factors, detailed $N$-body modelling will be required in order
to assess how significant evolutionary effects are on the $R_{\rm{h}}$ vs.
$M_V$ diagram, how they vary with cluster concentration, and whether they can 
explain the observed correlation.

\begin{figure}
\includegraphics[width=0.5\textwidth]{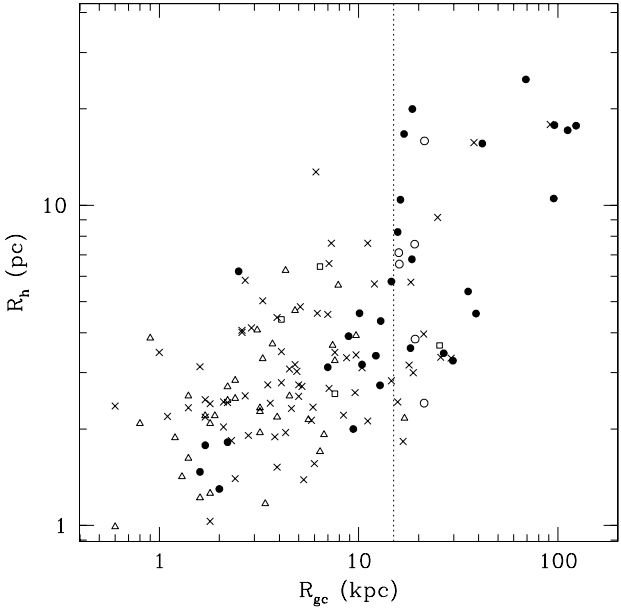}
\caption{Plot of $\log R_{\rm{h}}$ vs. $\log R_{\rm{gc}}$ for all Galactic globular
clusters. The BD clusters are marked by open triangles; the OH clusters
by crosses; the YH clusters by filled circles; the Sagittarius clusters
by open circles; and the unclassified (UN) clusters by open squares.
The vertical dotted line indicates $R_{\rm{gc}} = 15$ kpc.}
\label{f:rhrgc}
\end{figure}

An alternative possibility is that this correlation represents some
signature of globular cluster formation. If this is so, the details of this
signature are not immediately evident, as dynamical evolution (e.g., fading in $M_V$) 
must again be taken into account if the presently observed trend is to be traced 
back to its original form. In considering whether Eq. \ref{e:rhmv} is at least
partly due to cluster formation conditions, it is important to note that a 
number of previous authors (e.g., van den Bergh et al. \shortcite{vdb:91}; 
McLaughlin \shortcite{mclaughlin:00}) have claimed that a dependence of 
the half-light radius $R_{\rm{h}}$ on Galactocentric distance $R_{\rm{gc}}$ might 
contribute to the observed correlation between cluster size and luminosity 
for Galactic globular clusters. A physical interpretation of this argument 
is that the properties of the Galactic globular cluster system were set by 
the cluster formation process, which is environment-dependent -- that is, the 
parameters with which a cluster was formed were modulated by forces that depend on 
Galactocentric radius (see e.g., McLaughlin \shortcite{mclaughlin:00} 
for discussion). The basis for this idea can be seen in Figure \ref{f:rhrgc},
which shows a plot of $R_{\rm{h}}$ as a function of $R_{\rm{gc}}$ for the full 
sample of Galactic globular clusters. There is a clear trend of increasing 
$R_{\rm{h}}$ with increasing $R_{\rm{gc}}$. A formal fit to these data yields
\begin{equation}
\log R_{\rm{h}} = 0.42 \log R_{\rm{gc}} + 0.19,
\end{equation}
where the errors are $\pm 0.04$ on the both the slope and
intercept, and the correlation coefficient is $0.65$. This result is
consistent with those found previously by both McLaughlin 
\shortcite{mclaughlin:00} and van den Bergh et al. \shortcite{vdb:91}. 

In our present sample however, the extent to which a correlation between
$R_{\rm{h}}$ and Galactocentric radius can drive that between $R_{\rm{h}}$
and $M_V$ is not clear. All our Galactic clusters are in the outer halo,
with $R_{\rm{gc}} > 15$ kpc. Fig. \ref{f:rhrgc} shows that these objects
have a very large scatter in $R_{\rm{h}}$, and furthermore that diffuse clusters 
with $R_{\rm{gc}} \sim 15$ kpc do {\em not} have systematically smaller $R_{\rm{h}}$
than diffuse clusters with $R_{\rm{gc}} \sim 100$ kpc. Any apparent
trend in $R_{\rm{h}}$ with $R_{\rm{gc}}$ for these outer halo clusters
is most likely caused by the surprising {\em lack} of compact clusters at
very large Galactocentric radii (see Section \ref{ss:comphalo} below). 
In addition, the majority of the sample with $R_{\rm{gc}} > 15$ kpc are
YH members, and as such are very possibly accreted objects. In this
scenario, such clusters would not be expected to exhibit any trend 
between $R_{\rm{h}}$ and $R_{\rm{gc}}$ amongst themselves. We also note
that if the data in Fig. \ref{f:rhmvouter} are split into the Galactic and 
``external'' clusters, then fitting these two groups separately reproduces 
Eq. \ref{e:rhmv} for each to within $1\sigma$, albeit both with lower significance 
(due to the smaller number of points in each group).
It is curious that the globular clusters in the LMC, SMC, and Fornax and
Sagittarius dSph galaxies follow such a similar trend in Fig. \ref{f:rhmvouter}
to the outer halo Galactic clusters, even though they are members of
significantly different galaxies and have evolved in different environments.
Even so, it is observed that in the LMC, the most diffuse clusters (Reticulum, 
ESO121-SC03, NGC 1841, 2257) all lie at large galactocentric radii, as does the very
extended cluster 1 in the Fornax dSph, so there seems to be at least a weak dependence
of $R_{\rm{h}}$ on radius in these galaxies.

Ultimately, the most important conclusion to draw from Figures \ref{f:rhmvouter}
and \ref{f:rhrgc} is that the Galactic globular clusters with large
$R_{\rm{h}}$ are both faint, and occur at large Galactocentric radii.
With present data, it is not possible to identify for certain the origin 
of the observed trend in $R_{\rm{h}}$ with $M_V$. Although there is some
evidence that a correlation between $R_{\rm{h}}$ and $R_{\rm{gc}}$ could be
at least partly responsible, at present an intrinsic relationship between
$R_{\rm{h}}$ and $M_V$, perhaps a relic of the cluster formation process, cannot and
should not be ruled out. This puzzle seems well suited for study using large-scale 
realistic $N$-body simulations (see e.g., Aarseth \shortcite{aarseth:99}), which 
with the advent of powerful dedicated hardware (e.g., GRAPE-6 -- see Makino et 
al. \shortcite{makino:03}) are now nearly capable of direct globular cluster modelling.

\subsection{Where are the compact outer halo globular clusters?}
\label{ss:comphalo}
As noted above, examining Fig. \ref{f:rhrgc} in more detail reveals a surprising 
result: no compact YH clusters occur at $R_{\rm{gc}} > 40$ kpc.
This observation is intriguing because the presently accepted paradigm
is that the young halo objects have been accreted by the Galaxy via
minor mergers. Their $R_{\rm{h}}$ values should therefore be independent of 
$R_{\rm{gc}}$. This is so because the chaotic acquisition of clusters is expected 
to result in a wide spread of cluster sizes across a broad range of $R_{\rm{gc}}$
values. Indeed, this is exactly what is seen at intermediate
Galactocentric radii ($10 < R_{\rm{gc}} < 40$ kpc). The dispersion in $R_{\rm{h}}$ for 
the YH clusters at such radii is considerably larger than that for the 
clusters that are intrinsically associated with the Milky Way, and in fact 
closely matches the spread in $R_{\rm{h}}$ among the Sagittarius clusters. At very 
small $R_{\rm{gc}}$ we only really see compact YH clusters. However, recalling 
the apparent influence tidal forces have had on the OH and BD samples, this 
is not unexpected because extended clusters are especially vulnerable to rapid 
tidal destruction in the inner Galaxy.

It is therefore the handful of YH clusters at very large
Galactocentric radii\footnote{AM-1, Eridanus, Pyxis, Pal. 3, Pal. 4,
and Pal. 14} ($R_{\rm{gc}} > 40$ kpc) which are the most intriguing. A simple
examination of Fig. \ref{f:rhdist} shows that, among a randomly selected 
sample of six YH clusters, one should expect approximately three to have 
$R_{\rm{h}} < 5$ pc, or more than four to have $R_{\rm{h}} < 10$ pc. However, Fig. 
\ref{f:rhrgc} shows that there are {\em no} compact clusters among the six 
observed at very large Galactocentric radii. If the very outer young halo 
clusters originally belonged to now-defunct dwarf galaxies, as the presently 
accepted paradigm would seem to suggest, then there is no clear reason why 
their dispersion in $R_{\rm{h}}$ should not be similar to that observed for YH 
clusters with $10 < R_{\rm{gc}} < 40$ kpc.
   
\begin{figure}
\includegraphics[width=0.5\textwidth]{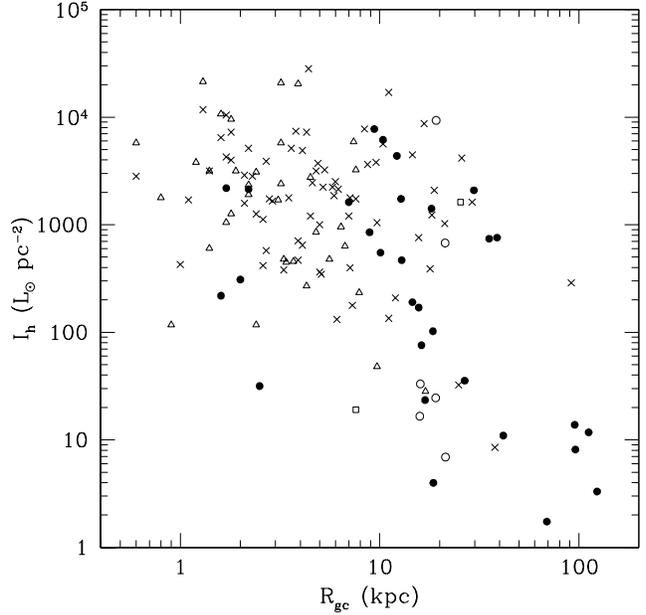}
\caption{Plot of $\log I_{\rm{h}}$ vs. $\log R_{\rm{gc}}$ for all Galactic globular
clusters. The point styles are as in Fig. \ref{f:rhrgc}.}
\label{f:sbrgc}
\end{figure}

Inspection of Table \ref{t:alldata} reveals an additional interesting fact. 
All of these six outer YH clusters are of relatively low luminosity (indeed, 
this result might be expected given Fig. \ref{f:rhmvouter} and Eq. 
\ref{e:rhmv}). The surface brightness $I_{\rm{h}}$ is derived from the 
parameters $R_{\rm{h}}$ and $M_V$, and is therefore useful in the present analysis. 
Fig. \ref{f:sbrgc} shows a plot of $I_{\rm{h}}$ as a function of 
$R_{\rm{gc}}$ for all Galactic globular clusters. In the inner Galaxy, there is no 
significant trend in $I_{\rm{h}}$ with $R_{\rm{gc}}$, but (as discussed in Section 
\ref{ss:sbdist}) there is a marked lack of low surface brightness 
clusters -- which are the most difficult to find against the backdrop of the 
Galactic bulge, and which are most vulnerable to tidal disruption. At 
intermediate $R_{\rm{gc}}$, the YH clusters are seen to have a significantly 
larger spread in $I_{\rm{h}}$ than do the native Galactic clusters. In fact, 
the dispersion in $I_{\rm{h}}$ observed at these Galactocentric radii for the 
YH clusters matches closely the spread in $I_{\rm{h}}$ exhibited by the 
Sagittarius objects. At $R_{\rm{gc}} > 40$ kpc only very low surface brightness 
clusters are found, i.e. only low luminosity clusters with large $R_{\rm{h}}$. Apart 
from NGC 2419, which is also interesting for other reasons (see Section 
\ref{ss:cores}, below), {\em there are no globular
clusters of even average surface brightness in the outer Galactic halo.}
   
\renewcommand{\labelenumi}{(\arabic{enumi})}
\renewcommand{\labelenumii}{(\alph{enumii})}

We can think of several possible scenarios that might account for
this lack of compact/luminous clusters at very large $R_{\rm{gc}}$:
\begin{enumerate}
\item{Clusters with $R_{\rm{gc}} > 40$ kpc once belonged to one, or more likely
several, now-destroyed parent dwarf galaxies which disintegrated at
large $R_{\rm{gc}}$ and which {\em only} possessed diffuse clusters at the 
time of merger.}
\item{Clusters with $R_{\rm{gc}} > 40$ kpc were outlying objects in one, or more
likely several, parent dwarf galaxies, and were lost very early
on during merger events which subsequently proceeded at smaller $R_{\rm{gc}}$.}
\item{Clusters with $R_{\rm{gc}} > 40$ kpc never belonged to a parent dwarf galaxy
and were formed as isolated objects at large Galactocentric radii.}
\end{enumerate}

Since little is known about the properties of galaxies which might have merged with
ours at early epochs, it is not possible to definitively rule out any of these
options. {\em If} these galaxies were similar to the Fornax and Sagittarius dSph
galaxies observed today, then options 1 and 2 seem implausible.
One could perhaps imagine a scenario in which any compact cluster members of 
a dwarf galaxy were formed close to the centre of the parent galaxy and soon 
merged into its core, leaving only diffuse clusters. However, both the Fornax and
Sagittarius dwarf galaxies contain clusters with a wide range of $R_{\rm{h}}$, so,
unless the hypothetical parent dwarf galaxies of the outer halo globulars
were radically different to Fornax and Sagittarius, option 1 appears
unlikely. In addition, Fornax is at a distance of $R_{\rm{gc}} \sim 140$ kpc 
and shows no signs of being strongly tidally influenced by the Galaxy, let 
alone disrupted. It also possesses two outlying clusters (Fornax 1 and 5) 
which have not been removed. Furthermore, the Sagittarius dwarf is 
{\em much} closer to the Galactic centre ($R_{\rm{gc}} \sim 20$ kpc), and still 
apparently possesses four clusters, of which two are diffuse objects 
(Terzan 8 and Arp 2). Hence, option 2 also appears unlikely in this scenario.

Having outlined this argument, we point out that we cannot be certain that
the galaxies which merged with ours at early epochs were in any way similar
to Fornax and Sagittarius. Indeed, studies of the chemical abundance patterns
in halo stars and in local dSph galaxies suggest that perhaps they were not
(see Mackey \& Gilmore \shortcite{mackey:04} for a brief summary). 
It is possible that the hypothetical parent dwarf galaxies of the outer YH
clusters, along with their member clusters, were systematically more diffuse 
than Fornax and Sagittarius and their clusters. If this were the case, it
might explain why these diffuse galaxies are now destroyed whereas Fornax and
Sagittarius are still (mostly) intact. In this scenario, we cannot rule out
options 1 and 2.
  
Option 3 also seems plausible, as the outermost YH clusters might help form
the upper end of a trend in increasing $R_{\rm{h}}$ with increasing $R_{\rm{gc}}$ 
if considered together with the BD and OH clusters (Fig. \ref{f:rhrgc}).
However, at least four of these outer young halo clusters are believed to be 
several Gyr younger than the oldest Galactic globulars (e.g., Stetson et al. 
\shortcite{stetson:99}; Sarajedini \shortcite{sarajedini:97}). This raises the 
question: why (and how) did cluster formation occur at such remote locations
{\em after} it occurred deeper within the Galactic potential well.
Furthermore, it is not clear why (or how) such cluster formation would
produce only diffuse, low luminosity clusters, and why these objects are
similar in so many ways to those found in local dwarf galaxies.

The outermost clusters are clearly very important tracers of how
remote parts of the Galactic halo were formed or assembled, as well as
the properties of the systems in which they might have formed. It is
therefore clearly of great interest to study them in considerably more detail
than has been achieved to date.

\subsection{The structures of very luminous clusters}
\label{ss:cores}

\begin{figure}
\includegraphics[width=0.5\textwidth]{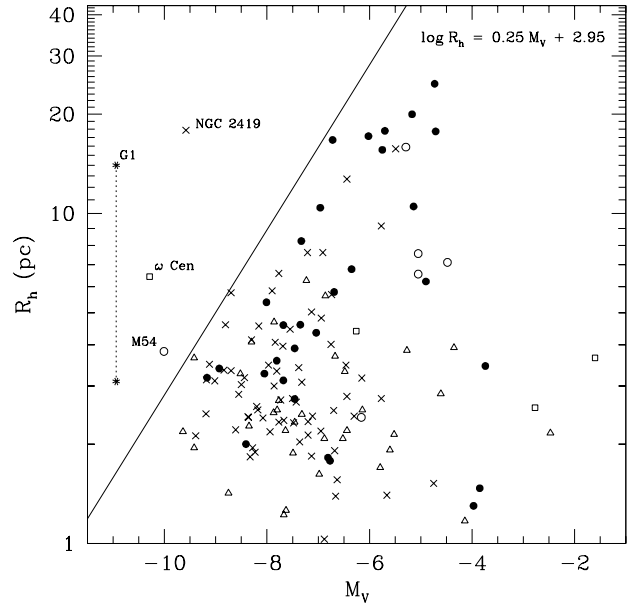}
\caption{Plot of $\log R_{\rm{h}}$ vs. $M_V$ for all Galactic globular
clusters. The point styles are again as in Fig. \ref{f:rhrgc}. The giant M31
globular cluster G1 is marked by two linked stars -- these represent the two 
discrepant measurements of $R_{\rm{h}}$ for this cluster. The upper point
is the measurement of Meylan et al. \shortcite{meylan:01}, while the lower
point is that of Barmby et al. \shortcite{barmby:02}.}
\label{f:rhmvall}
\end{figure}

Figure \ref{f:rhmvall} shows a plot of $\log R_{\rm{h}}$ vs. $M_V$ for all Galactic
globular clusters. The most important thing to note from this Figure is
that the upper left envelope of the main distribution of globular clusters
has a rather sharp edge. Only a very few clusters lie a significant
distance above this upper edge. If we shift the line defined by Eq. 
\ref{e:rhmv} upwards so that it has intercept $2.95$ (as plotted on Fig. 
\ref{f:rhmvall}) then one finds that the line
\begin{equation}
\log R_{\rm{h}} = 0.25 M_V + 2.95
\label{e:upper}
\end{equation}
closely describes the upper envelope of the cluster distribution. Only
three Galactic globular clusters lie above this line: $\omega$
Centauri, M54, and NGC 2419. A number of authors, such as Bekki \&
Freeman \shortcite{bekki:03}, Tsuchiya et al. \shortcite{tsuchiya:03}, and 
Ideta \& Makino \shortcite{ideta:04} (and references therein), have 
demonstrated that the observed properties of $\omega$ Centauri are 
consistent with it being the stripped core of a former dwarf spheroidal 
or dwarf elliptical galaxy. Furthermore, Ibata, Gilmore \& Irwin 
\shortcite{ibata:94,ibata:95} have shown that M54 (= NGC 6715) is
associated with the central region of the Sagittarius dwarf galaxy, and
some authors (see e.g. Layden \& Sarajedini \shortcite{layden:00}) have 
argued that M54 may actually be the core of this galaxy. Finally, in 
Paper I we suggested that NGC 2419 might also be the remaining core of a 
defunct dwarf spheroidal-type galaxy. In this connection it would clearly 
be of great interest to check if NGC 2419, like $\omega$ Centauri 
(e.g., Bedin et al. \shortcite{bedin:04}; Hilker et al. \shortcite{hilker:04}; 
and references therein), also shows evidence for age/metallicity substructure 
in its color-magnitude diagram. However, evidence for such substructure in 
NGC 2419 might be difficult to obtain because the mean metallicity 
($\langle[$Fe$/$H$]\rangle = -2.12$) of NGC 2419 is so much
lower than that of $\omega$ Centauri ($\langle[$Fe$/$H$]\rangle = -1.62$). 
As a result the expected range in metallicities for the population components 
of NGC 2419 is expected to be quite small. NGC 2419 is also significantly 
more distant than $\omega$ Centauri ($R_\odot = 84.2$ kpc), further 
complicating any observations.
   
It is intriguing that $\omega$ Centauri, M54, and NGC 2419, which have
all independently been previously suggested as the cores of
now-disrupted dwarf galaxies, should constitute a group of objects
which have such strikingly different structures than other luminous
Galactic clusters. This observation suggests a similar origin for these
three objects, since each has clearly evolved under quite different
circumstances and in different environments. Inspection of Table 
\ref{t:alldata} shows that three out of the four Galactic globular clusters 
brighter than $M_V = -9.5$ fall above Eq. \ref{e:upper}. The lone 
exception is NGC 6441 ($M_V = -9.64$, $R_{\rm{h}} = 2.2$ pc). The high 
metallicity ($[$Fe$/$H$] = -0.53$) of this cluster, which is located at 
$R_{\rm{gc}} = 3.9$ kpc, militates in favor of its assignment to the 
population associated with the Galactic bulge or inner disk. It
is noted in passing that the apparently normal globular NGC 5053 also
falls very close to (but still marginally below) the line defined by
Eq. \ref{e:upper}. 
   
Since the majority of the most luminous Galactic globular clusters
are now suspected of being the cores of former dwarf galaxies it seems
likely that the same might also be true in the globular cluster system
associated with M31. This suspicion is strengthened by the observation
that the cluster M31 Mayall No. II (= G1), which may be the most
luminous globular cluster associated with the Andromeda galaxy, has 
$M_V = -10.94$ and $R_{\rm{h}} = 14$ pc \cite{meylan:01}, although Barmby, 
Holland, \& Huchra \shortcite{barmby:02} find $R_{\rm{h}} = 3.1$ pc (using 
the distance of Meylan et al. \shortcite{meylan:01}). Adopting either of these 
$R_{\rm{h}}$ measurements, G1 also falls above the line defined 
by Eq. \ref{e:upper}, i.e. in the same region as $\omega$ Cen, M54, and 
NGC 2419. In this connection, we note that Meylan et al. \shortcite{meylan:01}
found their HST/WFPC2 photometry of G1 to be consistent with an
internal metallicity spread of $\sim 0.5$ dex. Furthermore
Bekki \& Chiba \shortcite{bekki:04} 
have recently argued that the observed properties of G1 can be 
explained if this object is the tidally stripped nucleus of a former 
dwarf elliptical (dE,N) galaxy.
   
Finally, Harris et al. \shortcite{harris:02} constructed a catalogue
of structural measurements for globular clusters in NGC 5128
(= Centaurus A). The $14$ most luminous of these have recently been
studied spectroscopically by Martini \& Ho \shortcite{martini:04},
who hypothesize that some of these clusters are actually the stripped nuclei 
of dwarf galaxies, based on their large masses and the possible detection
of extra-tidal light by Harris et al. \shortcite{harris:02} for several.
On the assumption that our line Eq. \ref{e:upper} provides a reasonable
means of discriminating objects which may be the cores of former dwarf
galaxies, the speculation of Martini \& Ho \shortcite{martini:04} is
supported by the observation that all $14$ of the clusters in their sample 
lie above and to the left of this line. In other words, all $14$ have structures 
and luminosities more similar to those of $\omega$ Cen, M54, NGC 2419, and 
G1 than to those of the majority of ``ordinary'' globular clusters 
associated with our Galaxy. Harris, Harris \& Geisler \shortcite{harris:04} 
have recently found that NGC 5128 contains $980 \pm 120$ globulars, so 
that potential dwarf galaxy remnants represent at least $\sim 1.4$
per cent of the clusters associated with this galaxy.

\section{Conclusions}
\label{s:conclusions}
In terms of their observed characteristics (metallicities, HB
morphologies, structures, shapes, luminosities, and surface
brightnesses) the Galactic globular clusters, considered as a whole,
constitute a relatively inhomogeneous class of objects. From a physical
point of view, it therefore makes sense to try to identify sub-
groups made up of clusters with similar characteristics. One of
the most successful classification schemes for the Galactic globular
clusters is that introduced by Zinn \shortcite{zinn:93} and 
van den Bergh \shortcite{vdb:93} (see also the references therein), 
whereby clusters are segregated according to their metallicity and 
HB morphology. Recently this scheme has been updated to cover the 
entire Galactic globular cluster system by Mackey \& Gilmore 
\shortcite{mackey:04}.
   
In the present paper, we have presented the complete list of cluster
classifications of Mackey \& Gilmore \shortcite{mackey:04}. Several 
objects in the Galactic halo possess properties which are consistent with 
the hypothesis that they are the stripped cores of former dwarf galaxies
(see below). The remaining majority of ``normal'' globular clusters fall
into three major categories -- bulge/disk (BD), ``old'' halo (OH), and
``young'' halo (YH) clusters. As recently re-emphasized by Mackey \&
Gilmore \shortcite{mackey:04} (see also Section \ref{s:properties}, above), 
the majority of BD and OH objects seem to have formed in the (proto) Galaxy, 
while the YH objects possess characteristics which indicate that they most 
likely formed in extra-Galactic systems and were accreted into the Galactic 
halo during subsequent merger events.
   
We have examined several of the properties of these three sub-systems
which were not considered by Mackey \& Gilmore \shortcite{mackey:04}. 
The BD, OH, and YH clusters typically occupy different regions of the 
Galaxy. The metal-rich BD objects are the most centrally concentrated, 
exhibiting a sharp cut-off in their Galactocentric radial distribution 
at $R_{\rm{gc}} = 7.9$ kpc. In contrast, the OH clusters dominate the inner 
halo, with the majority residing within $R_{\rm{gc}} = 20$ kpc, while the YH 
clusters typically occupy the outer halo, with more than half lying at 
$R_{\rm{gc}} > 15$ kpc.

In terms of cluster structures, luminosities, surface brightnesses,
and ellipticities, the YH clusters are clearly distinct from the OH and
BD objects. The OH and BD populations have apparently been strongly
modified by destructive tidal forces and bulge and disk shocks, so that
these sub-systems possess few extended clusters, few low luminosity
clusters, and few low surface brightness clusters. In comparison, the
YH group contains a significant fraction of extended clusters with
half-light radii $R_{\rm{h}} > 7$ pc, and low luminosity clusters with $M_V > -5.5$.
The luminosity distribution of YH clusters matches closely that observed
for clusters in four nearby dwarf galaxies (the LMC, SMC, and Fornax and
Sagittarius dwarf spheroidals), and as such, offers further evidence for
the external origin of much or all of the YH ensemble. By examining how
the observed distributions of half-light radii and mean surface
brightnesses for the YH clusters would have to be modified to match
those observed for native Galactic globulars, we estimate that the
original population of Galactic globular clusters may have numbered
$\sim 150-160$ objects. This result is consistent with several previous
estimates in the literature.
   
The distribution of cluster ellipticities shows that the YH clusters
are, on average, less flattened than are the OH and BD clusters. A
possible explanation of this result is that recent bulge and disk shocks
have introduced significant velocity anisotropy (and hence ellipticity)
into many BD and OH clusters \cite{goodwin:97}. We also note that all three
of the lowest surface brightness clusters for which ellipticity values
have been published are unusually flattened objects, as is the newly
discovered very low surface brightness object of Willman et al. 
\shortcite{willman:04}. This may indicate an intrinsic relationship 
between cluster rotation and surface brightness. An alternative, and 
perhaps more attractive, explanation invokes the influence of dark matter. 
In the globulars of lowest surface density the structure of the potential 
energy well might be dominated by a triaxial dark matter mini-halo 
\cite{mashchenko:04a,mashchenko:04b}, rather than by the baryonic material 
associated with Population II stars. The flattening of the isophotes of low 
surface brightness globular clusters might therefore provide information 
on their (still hypothetical) dark matter mini-halos. We briefly made the 
case for a comprehensive study of Galactic globular cluster ellipticities.
   
We have also investigated the relationship between cluster
luminosity and half-light radius. A strong correlation between the two
is observed for clusters in the outer Galactic halo and also among
clusters in the LMC, SMC, and Fornax and Sagittarius dwarf galaxies.
Several authors have previously asserted that any correlation between
$R_{\rm{h}}$ and $M_V$ for Galactic globular clusters can be explained by the fact
that $R_{\rm{h}}$ also appears to correlate with Galactocentric radius. However,
this cannot explain why the outer halo Galactic clusters, which are mostly YH
objects, follow the trend in $R_{\rm{h}}$ with $M_V$. If these objects are
accreted, no trend in $R_{\rm{h}}$ with $R_{\rm{gc}}$ is expected.

An additional puzzling observation is the lack of compact globular
clusters in the remote halo. We observe six YH clusters with $R_{\rm{gc}} > 40$
kpc; however all are diffuse and of low luminosity. Based on the
observed distributions of $R_{\rm{h}}$ and $M_V$ for YH clusters, one might have
expected several of the six to be compact and of greater than average
luminosity. We discussed several scenarios which might explain the
dearth of compact clusters with very large $R_{\rm{gc}}$; however none appear
particularly attractive.
   
Finally, if all the Galactic globular clusters are plotted on the
$\log R_{\rm{h}}$ vs. $M_V$ plane, several clusters stand well clear of the main 
body of objects. At very low luminosities we observe a number of compact
objects (Pal. 1, Pal. 13, E3 and AM-4) which have apparently
been strongly affected by shrinkage due to cooling and evaporation
of cluster stars, and stripping by Galactic tides.
The most interesting result, however, concerns the upper envelope of
the main distribution of clusters, which is sharply defined. Only three
Galactic globulars lie significantly above a line (Eq. \ref{e:upper}) 
describing this upper envelope -- $\omega$ Cen, M54, and NGC 2419. All 
three of these objects have previously, and independently, been hypothesized 
as being the stripped cores of former dwarf spheroidal-type galaxies. This 
suggests that the position of a cluster in the $\log R_{\rm{h}}$ vs. $M_V$ plane 
provides a powerful tool for the segregation between ``normal'' globular 
clusters and those which might previously have been associated with the cores
of dwarf galaxies. This suspicion is reinforced by the additional
observation that the massive cluster G1 in M31, which has been
previously suggested to be the remaining core of a nucleated dwarf
elliptical galaxy, also lies above Eq. \ref{e:upper}, as do $14$ of 
the most luminous clusters in NGC 5128, which have again been postulated 
as constituting the remnant cores of now defunct dwarf galaxies.

\section*{Acknowledgements}
We thank Peter Stetson for commenting on an early draft of this
paper, and George Wallerstein for correspondence about the color-
magnitude diagrams of dwarf spheroidals. We are also indebted to 
Marcin Sawicki, Liang Gao and Julio Navarro for helpful discussions, 
and we acknowledge the anonymous referee, whose thoughtful and constructive 
comments greatly improved this work. ADM recognises financial
support from PPARC in the form of a Postdoctoral Fellowship.



\bsp 

\label{lastpage}


\begin{thebibliography}{99}
\bibitem[\protect\citename{Aarseth }1999]{aarseth:99}
Aarseth S. J., 1999, Celest. Mech. Dynam. Astron., 73, 127
\bibitem[\protect\citename{Aarseth \& Heggie }1998]{aarseth:98}
Aarseth S. J., Heggie D. C., 1998, MNRAS, 297, 794
\bibitem[\protect\citename{Bailin et al.\ }2004]{bailin:04}
Bailin J., Steinmetz M., 2004, ApJ, submitted (astro-ph/0408163)
\bibitem[\protect\citename{Barmby et al.\ }2002]{barmby:02}
Barmby P., Holland S., Huchra J. P., 2002, AJ, 123, 1937
\bibitem[\protect\citename{Bedin et al.\ }2004]{bedin:04}
Bedin L. R., Piotto G., Anderson J., King I. R., Cassisi S., Momany Y.,
2004, Mem. Soc. Astr. It., in press (astro-ph/0406076)
\bibitem[\protect\citename{Bekki \& Freeman }2003]{bekki:03}
Bekki K., Freeman K. C., 2003, MNRAS, 346, L11
\bibitem[\protect\citename{Bekki \& Chiba }2004]{bekki:04}
Bekki K., Chiba M., 2004, A\&A, 417, 437
\bibitem[\protect\citename{C\^ot\'e }1999]{cote:99}
C\^ot\'e P., 1999, AJ, 118, 406
\bibitem[\protect\citename{C\^ot\'e et al.\ }2002]{cote:02}
C\^ot\'e P., Djorgovski S. G., Meylan G., Castro S., McCarthy J. K.,
2002, ApJ, 574, 783
\bibitem[\protect\citename{Davidge }2000]{davidge:00}
Davidge T. J., 2000, ApJS, 126, 105
\bibitem[\protect\citename{Dinescu et al.\ }1999]{dinescu:99}
Dinescu D. I., Girard T. M., van Altena W. F., 1999, AJ, 117, 1792
\bibitem[\protect\citename{Dinescu et al.\ }2003]{dinescu:03}
Dinescu D. I., Girard T. M., van Altena W. F., L\'opez C. E., 2003,
AJ, 125, 1373
\bibitem[\protect\citename{Fall \& Zhang }2001]{fall:01}
Fall S. M., Zhang Q., 2001, ApJ, 561, 751
\bibitem[\protect\citename{Frenk \& Fall }1982]{frenk:82}
Frenk C. S., Fall S. M., 1982, MNRAS, 199, 565
\bibitem[\protect\citename{Frinchaboy et al.\ }2004]{frinchaboy:04}
Frinchaboy P. M., Majewski S. R., Crane J. D., Reid I. N.,
Rocha-Pinto H. J., Phelps R. L., Patterson R. J., Mu\~noz R. R.,
2004, ApJ, 602, L21
\bibitem[\protect\citename{Gao et al.\ }2004]{gao:04}
Gao L., De Lucia G., White S. D. M., Jenkins A., 2004, MNRAS, 352, L1
\bibitem[\protect\citename{Goodwin }1997]{goodwin:97}
Goodwin S. P., 1997, MNRAS, 286, 39
\bibitem[\protect\citename{Gnedin \& Ostriker }1997]{gnedin:97}
Gnedin O. Y., Ostriker J. P., 1997, ApJ, 474, 223
\bibitem[\protect\citename{Harris et al.\ }2004]{harris:04}
Harris G. L. H., Harris W. E., Geisler D., 2004, AJ, 128, 723
\bibitem[\protect\citename{Harris }1996]{harris:96}
Harris W. E., 1996, AJ, 112, 1487
\bibitem[\protect\citename{Harris }2002]{harris:02}
Harris W. E., Harris G. L. H., Holland S. T., McLaughlin D. E., 2002,
AJ, 124, 1435 
\bibitem[\protect\citename{Helmi \& White }2001]{helmi:01}
Helmi A., White S. D. M., 2001, MNRAS, 323, 529
\bibitem[\protect\citename{Hilker et al.\ }2004]{hilker:04}
Hilker M., Kayser A., Richtler T., Wlllemsen P., 2004, A\&A, 422, 9
\bibitem[\protect\citename{Ibata et al.\ }1994]{ibata:94}
Ibata R. A., Gilmore G. F., Irwin M. J., 1994, Nature, 370, 194
\bibitem[\protect\citename{Ibata et al.\ }1995]{ibata:95}
Ibata R. A., Gilmore G. F., Irwin M. J., 1995, MNRAS, 277, 781
\bibitem[\protect\citename{Ibata et al.\ }2001]{ibata:01}
Ibata R. A., Lewis G. F., Irwin M., Totten E., Quinn T.,
2001, ApJ, 551, 294
\bibitem[\protect\citename{Ideta \& Makino }2004]{ideta:04}
Ideta M., Makino J., 2004, ApJ, in press (astro-ph/0408431)
\bibitem[\protect\citename{Inman \& Carney }1987]{inman:87}
Inman R. T., Carney B. W., 1987, AJ, 93, 1166
\bibitem[\protect\citename{Jiang \& Binney }2000]{jiang:00}
Jiang I. -G., Binney J., 2000, MNRAS, 314, 468
\bibitem[\protect\citename{Jord\'{a}n }2004]{jordan:04}
Jord\'{a}n A., 2004, ApJ, 613, L117
\bibitem[\protect\citename{Kanatas et al.\ }1995]{kanatas:95}
Kanatas I. N., Griffiths W. K., Dickens R. J., Penny A. J.,
1995, MNRAS, 272, 265
\bibitem[\protect\citename{Kavelaars \& Hanes }1997]{kavelaars:97}
Kavelaars J. J., Hanes D. A., 1997, MNRAS, 285, L31
\bibitem[\protect\citename{King }1962]{king:62}
King I., 1962, AJ, 67, 471
\bibitem[\protect\citename{Layden \& Sarajedini }2000]{layden:00}
Layden A. C., Sarajedini A., 2000, AJ, 119, 1760
\bibitem[\protect\citename{Lee }1990]{lee:90}
Lee Y. -W., 1990, ApJ, 363, 159
\bibitem[\protect\citename{Lee et al.\ }1994]{lee:94}
Lee Y. -W., Demarque P., Zinn R., 1994, ApJ, 423, 248
\bibitem[\protect\citename{Lightman \& Shapiro }1978]{lightman:78}
Lightman A. P., Shapiro S. L., 1978, Rev Mod. Phys. 50, 437
\bibitem[\protect\citename{Mackey \& Gilmore }2003a]{mackey:03a}
Mackey A. D., Gilmore G. F., 2003a, MNRAS, 340, 175
\bibitem[\protect\citename{Mackey \& Gilmore }2003b]{mackey:03b}
Mackey A. D., Gilmore G. F., 2003b, MNRAS, 343, 747
\bibitem[\protect\citename{Mackey \& Gilmore }2004]{mackey:04}
Mackey A. D., Gilmore G. F., 2004, MNRAS, in press
\bibitem[\protect\citename{Makino et al.\ }2003]{makino:03}
Makino J., Fukushige T., Koga M., Namura K., 2003, PASJ, 55, 1163
\bibitem[\protect\citename{Martini \& Ho }2004]{martini:04}
Martini P., Ho L. C., 2004, ApJ, in press (astro-ph/0404003)
\bibitem[\protect\citename{Mashchenko \& Sills }2004a]{mashchenko:04a}
Mashchenko S., Sills A., 2004a, ApJ, in press (astro-ph/0409605)
\bibitem[\protect\citename{Mashchenko \& Sills }2004b]{mashchenko:04b}
Mashchenko S., Sills A., 2004b, ApJ, in press (astro-ph/0409606)
\bibitem[\protect\citename{McClure et al.\ }1985]{mcclure:85}
McClure R. D., Hesser J. E., Stetson P. B., Stryker L. L.,
1985, PASP, 97, 665
\bibitem[\protect\citename{McLaughlin }2000]{mclaughlin:00}
McLaughlin D. E., 2000, ApJ, 539, 618
\bibitem[\protect\citename{Meylan et al.\ }2001]{meylan:01}
Meylan G., Sarajedini A., Jablonka P., Djorgovski S. G., 
Bridges T., Rich R. M., 2001, AJ, 122, 830
\bibitem[\protect\citename{Ortolani et al.\ }1997]{ortolani:97}
Ortolani S., Bica E., Barbuy B., 1997, MNRAS, 284, 692
\bibitem[\protect\citename{Ostriker \& Gnedin }1997]{ostriker:97}
Ostriker J. P., Gnedin O. Y., 1997, ApJ, 487, 667
\bibitem[\protect\citename{Piotto et al.\ }1997]{piotto:97}
Piotto G., Cool A. M., King I. R., 1997, AJ, 113, 1345
\bibitem[\protect\citename{Rey et al.\ }2001]{rey:01}
Rey S. -C., Yoon S. -J., Lee Y. -W., Chaboyer B., Sarajedini A.,
2001, AJ, 122, 3219
\bibitem[\protect\citename{Rosenberg et al.\ }1998]{rosenberg:98}
Rosenberg A., Saviane I., Piotto G., Aparicio A., Zaggia S. R.,
1998, AJ, 115, 648
\bibitem[\protect\citename{Salaris \& Weiss }2002]{salaris:02}
Salaris M., Weiss A., 2002, A\&A, 388, 492
\bibitem[\protect\citename{Sarajedini }1997]{sarajedini:97}
Sarajedini A., 1997, AJ, 113, 682
\bibitem[\protect\citename{Siegel et al.\ }2001]{siegel:01}
Siegel M. H., Majewski S. R., Cudworth K. M., Takamiya M.,
2001, AJ, 121, 935
\bibitem[\protect\citename{Spitzer \& Thuan }1972]{spitzer:72}
Spitzer L., Thuan T. X., 1972, ApJ, 175, 31
\bibitem[\protect\citename{Stetson et al.\ }1999]{stetson:99}
Stetson P. B., et al., 1999, AJ, 117, 247
\bibitem[\protect\citename{Tsuchiya et al.\ }2003]{tsuchiya:03}
Tsuchiya T., Dinescu D. I., Korchagin V. I., 2003, ApJ, 598, L29
\bibitem[\protect\citename{van den Bergh }1993]{vdb:93}
van den Bergh S., 1993, ApJ, 411, 178
\bibitem[\protect\citename{van den Bergh }1994]{vdb:94}
van den Bergh S., 1994, AJ, 108, 2145
\bibitem[\protect\citename{van den Bergh \& Morbey}1984]{vdb:84}
van den Bergh S., Morbey C. L., 1984, Astronomy Express, 1, 1
\bibitem[\protect\citename{van den Bergh \& Mackey }2004]{paper1}
van den Bergh S., Mackey A. D., 2004, MNRAS, 354, 713 (= Paper I)
\bibitem[\protect\citename{van den Bergh et al.\ }1980]{vdb:80}
van den Bergh S., Demers S., Kunkel W. E., 1980, ApJ, 239, 112
\bibitem[\protect\citename{van den Bergh et al.\ }1991]{vdb:91}
van den Bergh S., Morbey C., Pazder J., 1991, ApJ, 375, 594
\bibitem[\protect\citename{Webbink }1985]{webbink:85}
Webbink R. F., 1985, in Goodman J., Hut P., eds., Proc. IAU Symp. 113,
Dynamics of Star Clusters. Kluwer, Dordrecht, p. 541
\bibitem[\protect\citename{Willman et al.\ }2004]{willman:04}
Willman B., et al., 2004, AJ, submitted (astro-ph/0410416)
\bibitem[\protect\citename{Zinn }1993]{zinn:93}
Zinn R., 1993a, in Smith G. H., Brodie J. P., eds., ASP Conf. Ser. 48,
The Globular Cluster-Galaxy Connection. Astron. Soc. Pac.,
San Francisco, p. 38
\end{thebibliography}
\end{document}